\documentclass[conference,compsoc]{IEEEtran}

\usepackage{subfigure}  
\usepackage{caption}
\usepackage{hegroup}
\usepackage{amsmath}
\usepackage{graphicx}
\usepackage{tabularx}
\usepackage{booktabs}
\usepackage{multirow}
\usepackage{amsfonts}
\usepackage{rotating}
\usepackage{array}
\usepackage{algorithm}
\usepackage{comment}
\usepackage{todonotes}
\usepackage{xcolor}
\usepackage{dsfont}  

\usepackage[noEnd=True]{algpseudocodex}

\usepackage{tcolorbox}
\usepackage{cleveref}
\usepackage{verbatim} 
\usepackage{enumitem}

\usepackage[edges]{forest}
\definecolor{hidden-draw}{RGB}{106,142,189} 
\definecolor{hidden-blue}{RGB}{194,232,247} 
\definecolor{hidden-orange}{RGB}{217, 232, 252} 
\begin{document}

\title{SoK: a Comprehensive Causality Analysis Framework for Large Language Model Security}

\author{Wei Zhao\textsuperscript{*}\thanks{* These authors contributed to the work equally and should be regarded as co-first authors.}, Zhe Li\textsuperscript{*}, 
Jun Sun, \\
Singapore Management University\\
\texttt{\{wzhao,zheli,junsun\}@smu.edu.sg}\\
}

\date{}
\maketitle

\begin{abstract}
Large Language Models (LLMs) exhibit remarkable capabilities but remain vulnerable to adversarial manipulations such as jailbreaking, where crafted prompts bypass safety mechanisms. Understanding the causal factors behind such vulnerabilities is essential for building reliable defenses.

In this work, we introduce a unified causality analysis framework that systematically supports all levels of causal investigation in LLMs, ranging from token-level, neuron-level, and layer-level interventions to representation-level analysis. The framework enables consistent experimentation and comparison across diverse causality-based attack and defense methods. Accompanying this implementation, we provide the first comprehensive survey of causality-driven jailbreak studies and empirically evaluate the framework on multiple open-weight models and safety-critical benchmarks including jailbreaks, hallucination detection, backdoor identification, and fairness evaluation. Our results reveal that: (1) targeted interventions on causally critical components can reliably modify safety behavior; (2) safety-related mechanisms are highly localized (i.e., concentrated in early-to-middle layers with only 1–2\% of neurons exhibiting causal influence); and (3) causal features extracted from our framework achieve over 95\% detection accuracy across multiple threat types.

By bridging theoretical causality analysis and practical model safety, our framework establishes a reproducible foundation for research on causality-based attacks, interpretability, and robust attack detection and mitigation in LLMs. Code is available at \url{https://github.com/Amadeuszhao/SOK_Casuality}.
\end{abstract}


\section{Introduction}

Large Language Models (LLMs), such as ChatGPT~\cite{BMRSKDNSSAAHKHCRZWWHCSLGCCBMRSA20} and Gemini~\cite{team2023gemini}, have revolutionized natural language processing, achieving remarkable performance in tasks such as question answering~\cite{BMRSKDNSSAAHKHCRZWWHCSLGCCBMRSA20}, code completion~\cite{chen2021evaluating}, and text generation~\cite{brown2020language}. However, as these models are increasingly deployed in real-world applications, their vulnerabilities pose significant security and reliability risks. A particularly severe threat is \emph{jailbreaking}, i.e., the use of adversarial prompts to bypass safety mechanisms and elicit harmful or policy-violating outputs~\cite{Yao_2024,GAAPP23,SAN23}. Such attacks have broad implications, from enabling misinformation and privacy breaches to undermining the trustworthiness of AI-assisted decision systems.

While several surveys have cataloged jailbreak attacks and defenses~\cite{xu-etal-2024-comprehensive,yi2024jailbreakattacksdefenseslarge}, existing work has primarily taken a descriptive approach, lacking systematic tools to analyze \emph{why} such vulnerabilities arise. In this work, we advance a causality-centered perspective: we argue that the key to understanding and mitigating jailbreaks lies in uncovering the \emph{causal mechanisms} within LLMs that govern their safety behavior. Rather than treating jailbreaks as black-box failures, we analyze how interventions at multiple levels (including tokens, neurons, layers, and representations) propagate through the model to shape outputs.

To this end, we present a unified causality analysis framework that supports comprehensive, multi-level causal investigation of LLMs. The framework provides an extensible platform for implementing and comparing causality-based attacks and defenses, enabling systematic study of how causal dependencies within transformer architectures influence model safety. Specifically, it integrates four analytical levels:
\begin{itemize}
    \item \emph{Token-level analysis}, which examines how input tokens causally affect outputs via counterfactual or replacement interventions;
    \item \emph{Neuron-level analysis}, which identifies sparse, causally critical neurons that modulate harmful or safe behavior;
    \item \emph{Layer-level analysis}, which traces how causal influence propagates through transformer layers to shape safety-related decisions; and
    \item \emph{Representation-level analysis}, which explores how embedding geometry encodes safety boundaries and how attacks perturb these structures.
\end{itemize}

Building upon this framework, we also conduct the \emph{first comprehensive survey} of causality-based jailbreak research and \emph{empirically evaluate} the framework’s effectiveness across multiple open-weight models, including Llama2-7B, Qwen2.5-7B, and Llama3.1-8B, on diverse safety-critical benchmarks including jailbreaks, hallucination detection, backdoor identification, and fairness evaluation.

Our experiments yield three central findings. First, targeted interventions on causally critical components identified by our framework reliably alter model safety behavior, demonstrating causal leverage. Second, causal localization studies reveal that safety mechanisms are concentrated in early-to-middle transformer layers (2–12), with only 1–2\% of neurons exhibiting safety-critical roles. Third, causal features extracted from our multi-level analysis enable highly effective detection and defense, achieving detection success rates above 95\% across jailbreak, hallucination, backdoor, and fairness tasks.

Together, these results establish our framework as a general foundation for causal analysis of LLM vulnerabilities. By connecting mechanistic insights with practical security evaluation, this work reframes jailbreak research as both a security problem and a scientific opportunity to uncover the causal structures that underlie LLM safety.

Our main contributions are as follows:
\begin{itemize}
    \item A unified framework for causality analysis across token, neuron, layer, and representation levels, supporting reproducible and extensible evaluation of diverse causality-based attacks and defenses;
    \item A comprehensive survey of causality-driven jailbreak research, situating prior work within a consistent multi-level causal taxonomy; and
    \item A systematic empirical evaluation demonstrating how causal interventions and features derived from the framework inform both model understanding and robust defense strategies.
\end{itemize}

By centering causality as the unifying lens, this work bridges theoretical understanding and practical security, offering a principled pathway toward more interpretable, controllable, and trustworthy large language models. Our framework is publicly available at 
\url{https://github.com/Amadeuszhao/SOK_Casuality}.

\section{Background} \label{sec: background}
In this section, we survey existing jailbreak attacks and corresponding defense mechanisms, with particular emphasis on causality-based approaches. While our broader causality-analysis framework offers deeper mechanistic insights into model behavior across a range of settings, we center the discussion on jailbreak attacks, as they provide a concrete, intuitive, and extensively studied use case.

\tikzstyle{my-box}=[
 rectangle,
 draw=hidden-draw,
 rounded corners,
 text opacity=1,
 minimum height=1.5em,
 inner sep=2pt,
 align=center,
 fill opacity=.5,
 ]
 \tikzstyle{leaf}=[my-box, minimum height=1.5em,
 fill=hidden-orange!60, text=black, align=center,font=\scriptsize,
 inner xsep=2pt,
 inner ysep=4pt,
 ]

\begin{figure*}[!ht]
    \centering
    \resizebox{0.75\textwidth}{!}{
        \begin{forest}
            forked edges,
            for tree={
                grow=east,
                reversed=true,
                anchor=base west,
                parent anchor=east,
                child anchor=west,
                node options={align=center},
                align=center,
                base=left,
                font=\small,
                rectangle,
                draw=hidden-draw,
                rounded corners,
                edge+={darkgray, line width=1pt},
                s sep=3pt,
                inner xsep=2pt,
                inner ysep=3pt,
                ver/.style={rotate=90, child anchor=north, parent anchor=south, anchor=center},
            },
            where level=1{text width=6.5em,font=\scriptsize}{},
            where level=2{text width=6.5em,font=\scriptsize}{},
            where level=3{text width=6.5em,font=\scriptsize}{},
            where level=4{text width=6.5em,font=\scriptsize}{},
            [Jailbreak Attack Methods, ver
                [Automated \\ Attacks
                    [Optimization-Based \\ Attacks
                        [GCG \cite{GCG23} \\ IGCG \cite{IGCG24} \\ PRP \cite{2024prp} \\ RLbreaker \cite{RLbreaker24}, leaf]
                    ]
                    [Iterative \\ Refinement Attacks
                        [JSAA \cite{JSAA24} \\ PAIR \cite{PAIR23} \\ TAP \cite{TAP23} \\ LLM-Fuzzer \cite{LLM-Fuzzer24} \\ DeepInception \cite{li2023deepinception} \\ AutoDAN \cite{AutoDAN24} \\ AutoDan-Turbo \cite{autodanturbo25} \\ Cipher \cite{cipher2024}, leaf]
                    ]
                     [Generator-Based \\ Attacks
                            [MASTERKEY \cite{MASTERKEY24} \\ Advprompter \cite{Advprompter24} \\ AmpleGCG \cite{amplegcg2024}, leaf]
                        ]
                ]
                [\textcolor{blue}{Causality-Guided} \\ \textcolor{blue}{Attacks}
                    [\textcolor{blue}{Training Analysis} 
                        [\textcolor{blue}{Jailbroken} \cite{Jailbroken23} \\ \textcolor{blue}{IFSJ} \cite{IFSJ24} \\ \textcolor{blue}{GEA} \cite{GEA24}, leaf]
                    ]
                    [\textcolor{blue}{Component Analysis} 
                        [\textcolor{blue}{Rep Steering} \cite{2024understanding} \\ \textcolor{blue}{NeuroStrike} \cite{2025neurostrike} \\ \textcolor{blue}{Casper} \cite{zhao2023causality}, leaf]
                    ]
                ]
            ]
        \end{forest}
    }
    \caption{Taxonomy of jailbreak attack methods with emphasis on causality-guided approaches.}
    \label{fig: Attack}
\end{figure*}
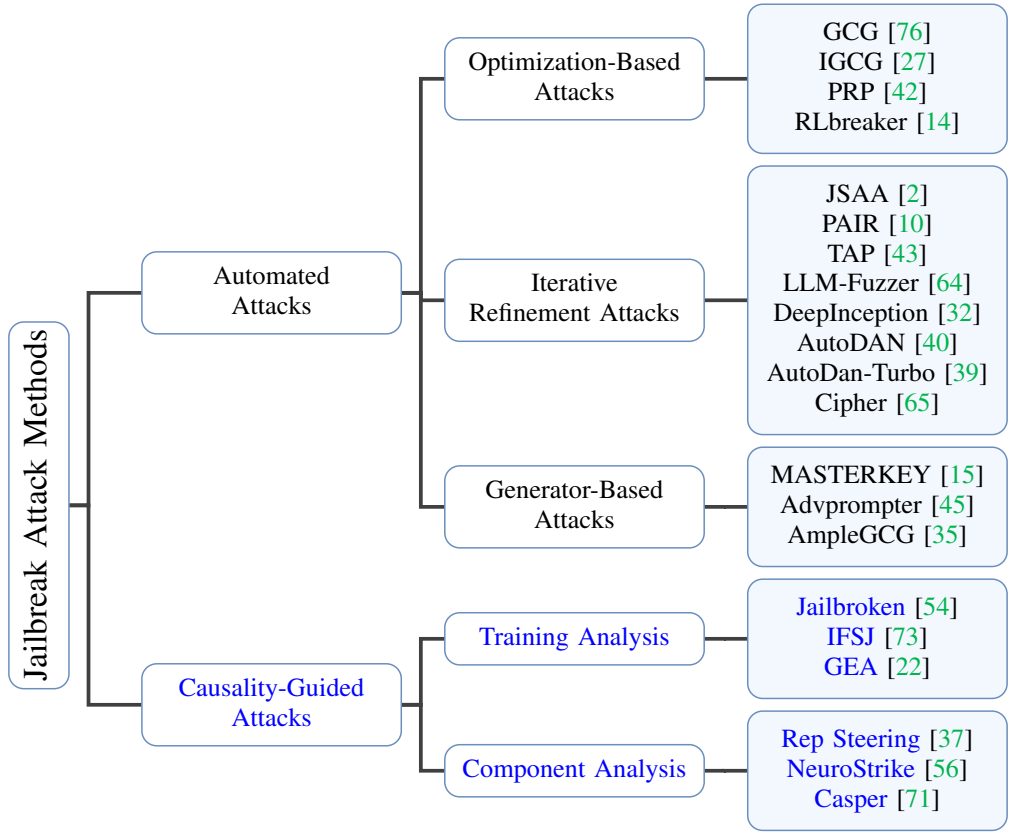
\subsection{Jailbreak Attacks}
Jailbreak attacks are adversarial prompting techniques that intentionally bypass the safety alignment of LLMs, compelling them to produce restricted or harmful outputs. We organize existing approaches into two principal categories, as shown in Figure~\ref{fig: Attack}, i.e., Automated Attacks (that are not based on causality) and Causality-Guided Attacks. While the former explores various strategies to craft adversarial prompts through optimization, iterative refinement, or direct generation from fine-tuned models, our primary focus is on causality-guided attacks, which exploit the underlying causal mechanisms that drive alignment behavior. \\

\noindent\textbf{Automated Attacks.}
 Automated attacks treat jailbreak construction as a computational problem, systematically generating adversarial prompts through algorithmic approaches. Optimization-based attacks~\cite{GCG23, IGCG24, 2024prp,RLbreaker24} leverage gradient signals or search algorithms to identify token substitutions that maximize harmful output probability, achieving high success rates and cross-model transferability despite requiring significant computational resources. Iterative refinement attacks~\cite{JSAA24, PAIR23, TAP23, LLM-Fuzzer24, li2023deepinception, AutoDAN24, autodanturbo25, cipher2024} leverage LLMs themselves to propose, test, and refine candidate prompts through feedback-driven loops or evolutionary strategies, producing human-readable adversarial prompts that adapt efficiently with fewer queries. While both approaches require multiple time-consuming steps to construct each jailbreak prompt, generator-based attacks~\cite{MASTERKEY24,amplegcg2024,Advprompter24} circumvent this limitation by fine-tuning models on successful jailbreak examples along with their mutations and variations. Once trained, these generators can reliably produce diverse adversarial prompts in a single forward pass without further optimization, facilitating efficient large-scale attacks. \\

\noindent\textbf{Causality-Guided Attacks.} Causality-guided attacks enhance jailbreak effectiveness by revealing the underlying mechanisms that make manipulations successful, enabling minimal and precise interventions that reliably override safety decisions. Unlike prompt engineering techniques which exhibit inconsistent performance, these attacks target the underlying decision-making processes themselves through rigorous analysis of internal model components or training procedures.

\emph{Training Process Analysis.} Training-level causality analysis examines how different training objectives causally influence model behavior to uncover fundamental weaknesses in safety mechanisms. The Jailbroken framework~\cite{Jailbroken23} identifies two failure modes: competing training objectives that create potential conflicts between pretraining, instruction-following, and safety alignment (exploited through prefix injection and refusal suppression), and mismatched generalization where pretraining extends to domains not adequately covered by safety training (exploited through obfuscation strategies like Base64 encoding or cross-lingual prompts). Building on these insights, I-FSJ~\cite{IFSJ24} and GEA~\cite{GEA24} demonstrate that simple manipulations, whether through in-context learning with harmful demonstrations and special tokens or through generation setting adjustments that remove safety prompts and alter decoding parameters, can effectively compromise safety alignment with significantly reduced computational costs.

\tikzstyle{my-box}=[
 rectangle,
 draw=hidden-draw,
 rounded corners,
 text opacity=1,
 minimum height=1.5em,
 inner sep=2pt,
 align=center,
 fill opacity=.5,
 ]
 \tikzstyle{leaf}=[my-box, minimum height=1.5em,
 fill=hidden-orange!60, text=black, align=center,font=\scriptsize,
 inner xsep=2pt,
 inner ysep=4pt,
 ]
\begin{figure*}[t]
    \centering
    \resizebox{0.8\textwidth}{!}{
        \begin{forest}
            forked edges,
            for tree={
                grow=east,
                reversed=true,
                anchor=base west,
                parent anchor=east,
                child anchor=west,
                node options={align=center},
                align=center,
                base=left,
                font=\small,
                rectangle,
                draw=hidden-draw,
                rounded corners,
                edge+={darkgray, line width=1pt},
                s sep=3pt,
                inner xsep=2pt,
                inner ysep=3pt,
                ver/.style={rotate=90, child anchor=north, parent anchor=south, anchor=center},
            },
            where level=1{text width=8.0em,font=\scriptsize}{},
            where level=2{text width=8.0em,font=\scriptsize}{},
            where level=3{text width=7.5em,font=\scriptsize}{},
            where level=4{text width=8.0em,font=\scriptsize}{},
            [Jailbreak Defense Methods, ver
                [Inference-based Defense
                    [Input-based Defense
                            [Perplexity \cite{Perplexity2308} \cite{jain2023baseline} \\  
                            IAPrompt \cite{IAPrompt2401} \\
                            Self-Reminder \cite{SelfReminder23} \\ RPO \cite{RPO2401} \\ ICD \cite{ICD24}, leaf, text width=7em]
                    ]
                    [Output-based Defense
                            [Self-Examination \cite{LLMSelfDefense23} \\ RALLLM \cite{RALLM2309} \\ SmoothLLM \cite{SmoothLLM2310} \cite{SemanticSmooth2402} \\ Llama Guard \cite{LlamaGuard2312} \\ SafeDecoding \cite{SafeDecoding2402}, leaf, text width=7em]
                    ]
                    [Process-based Defense
                            [GradSafe \cite{GradSafe2402} \\ Gradient Cuff \cite{GradCuff2403} \\
                            RAIN \cite{RAIN2309} \\ Goal Prioritization \cite{GoalPrioritization24}, leaf, text width=7em]
                    ]
                    [\textcolor{blue}{Causality-based Detection}
                        [\textcolor{blue}{Erase-and-Check} \cite{EraseCheck2309} \\
                        \textcolor{blue}{DRO} \cite{DRO24} \\ 
                        \textcolor{blue}{LLMScan} \cite{zhang2024llmscan}, leaf, text width=7em]
                    ]
                ]
                [Model-based Defense
                    [Adversarial Training
                        [CAT \cite{CAT24} \\ Adversarial Tuning \cite{AdversarialTuning2406} \\
                        Constitution AI \cite{constitutional22} \cite{BB2025} \\
                        Eraser \cite{Eraser24} \cite{UNlearn25} , leaf, text width=7.5em]
                    ]
                    [\textcolor{blue}{Causality-Guided Tuning}
                        [\textcolor{blue}{LED} \cite{LED2405} \\ \textcolor{blue}{Safe Neurons} \cite{2025neurons} \\
                        \textcolor{blue}{Circuit Breaking} \cite{CircuitBreaker24} \\ 
                        \textcolor{blue}{DELMAN} \cite{2025delman},  leaf, text width=7.5em]
                    ]
                ]
            ]
        \end{forest}
    }
    \caption{Taxonomy of jailbreak defense methods with emphasis on causality-based techniques.}
    \label{Defense}
\end{figure*}
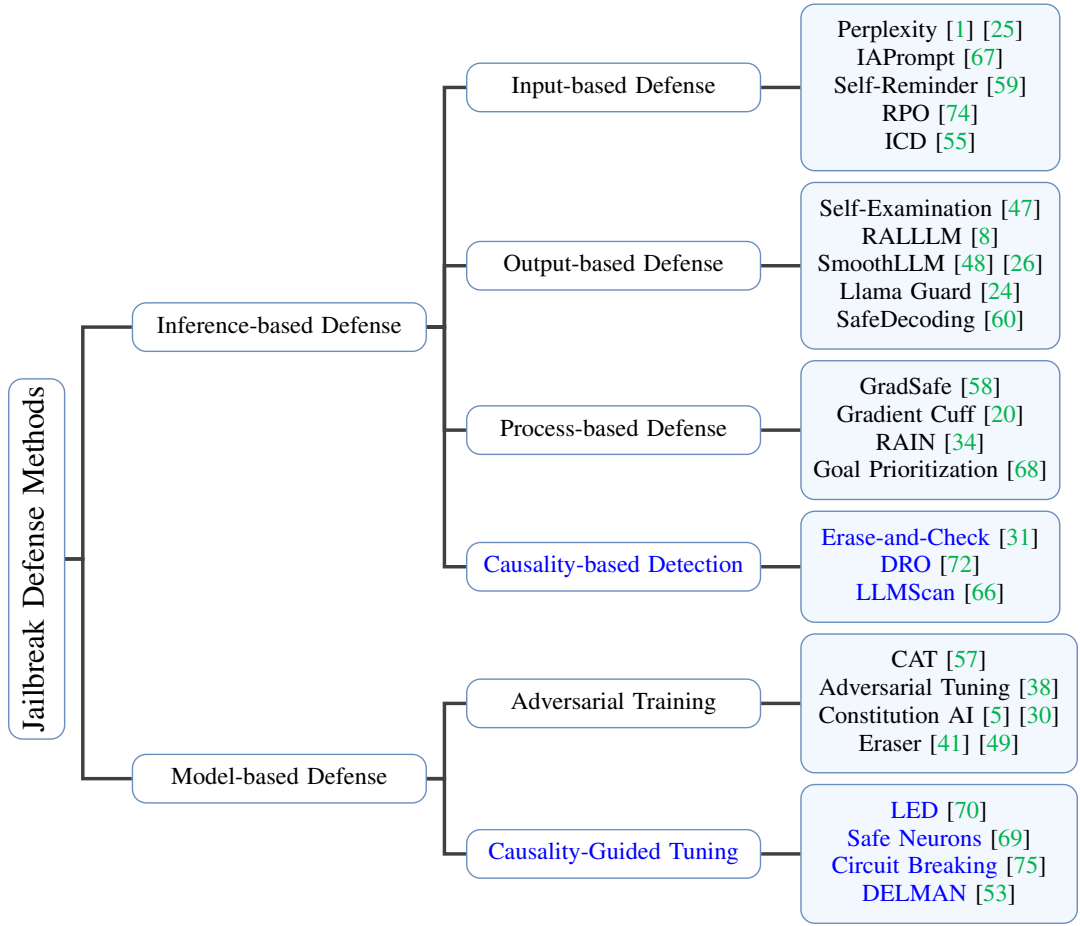

\emph{Component Analysis.} Component-level causality analysis identifies which internal model elements—layers, representations, or neurons—are causally responsible for safety outcomes. CASPER~\cite{zhao2023causality} conducts layer-based causality analysis through pruning interventions, revealing that safety alignment relies on overfitting in early layers that function as discriminators for harmful prompts. Their emoji attack exploits this by distributing causal influence more evenly across layers to bypass detection. Representation Steering~\cite{2024understanding} analyzes how jailbreak attacks manifest in LLM representation spaces, discovering that successful jailbreaks systematically move harmful prompt representations toward the "acceptance direction," the geometric direction leading to compliance rather than refusal, enabling optimization objectives that explicitly guide prompts along this direction. NeuroStrike~\cite{2025neurostrike} operates at neuron-level, identifying sparse "safety neurons" through logistic regression classifiers and z-score analysis, then selectively manipulating these neurons while demonstrating transferability across model variants.

\subsection{Jailbreak Defense}
The rise of sophisticated jailbreak techniques has heightened concerns over LLM safety, exposing significant risks to both commercial and open-source models. In response, the research community has developed a variety of defense strategies that can be broadly divided into two main categories based on their mode of intervention, as illustrated in Figure~\ref{Defense}: (1) Inference-level Defense and (2) Model-level Defense. This distinction highlights a fundamental trade-off between deployment flexibility and lasting safety improvements.

\noindent\textbf{Inference-level Defense.}
Inference-level defenses operate during the model's inference process without modifying the underlying model weights, providing crucial real-time protection against jailbreak attacks.

\textit{Input-based Defense.} These methods serve as the first line of protection by examining and modifying user prompts before they reach the model's generation process. Detection methods identify potentially harmful inputs through perplexity analysis~\cite{Perplexity2308,jain2023baseline} or intent analysis~\cite{IAPrompt2401}. Mitigation methods actively modify or augment prompts to neutralize harmful content, including Self-Reminder~\cite{SelfReminder23} and In-Context Defense (ICD)~\cite{ICD24} which incorporate safety instructions or demonstration examples, and Robust Prompt Optimization~\cite{RPO2401} which optimizes universal defensive suffixes through minimax optimization.

\textit{Output-based Defense.} These methods focus on evaluating and modifying generated responses to ensure safety compliance before delivery to users. Detection methods assess response safety through Self-Examination~\cite{LLMSelfDefense23}, dedicated safety classifiers like Llama-Guard~\cite{LlamaGuard2312}, or perturbation-driven approaches~\cite{RALLM2309,SmoothLLM2310,SemanticSmooth2402} that evaluate consistency across input variants. Mitigation methods such as SafeDecoding~\cite{SafeDecoding2402} modify token probability distributions during decoding to amplify safety-related tokens and steer generation toward safer outputs.

\textit{Process-based Defense.} These methods monitor and intervene during the generation process itself through continuous safety assessment. Detection methods analyze internal model states including gradient patterns of safety-critical parameters~\cite{GradSafe2402} or refusal loss properties~\cite{GradCuff2403}. Mitigation methods actively guide generation toward safer responses through iterative self-evaluation and rewinding mechanisms~\cite{RAIN2309} or priority balancing between helpfulness and safety~\cite{GoalPrioritization24}.

\textit{Causality-based Detection.} A growing body of research leverages causal analysis to identify the underlying causal signals that give rise to harmful behaviors during LLM inference. The Erase-and-check framework~\cite{EraseCheck2309} systematically removes tokens from input prompts and evaluates remaining subsequences using safety filters, providing fine-grained causal insight into which prompt components contribute to harmful intent. Directed Representation Optimization (DRO)~\cite{DRO24} investigates how safety prompts shape model responses through internal representation analysis, revealing that safety prompts shift query representations toward a "higher-refusal" direction. LLMScan~\cite{zhang2024llmscan} constructs detailed causal maps through systematic interventions at both token and layer levels, enabling fine-grained identification of harmful prompts by tracing how causal dependencies propagate throughout the model. \\

\noindent\textbf{Model-level Defense.}
Model-level defenses involve permanent modifications to language model parameters through training, fine-tuning, or direct parameter manipulation to enhance intrinsic safety capabilities.

\textit{Adversarial Training.} These methods systematically expose models to adversarial examples during training to improve robustness. Continuous Adversarial Training (CAT)~\cite{CAT24} operates in continuous embedding space to achieve efficient robustness against gradient-based attacks. Adversarial Tuning~\cite{AdversarialTuning2406} implements hierarchical meta-universal adversarial prompt learning for both token-level and semantic-level defense. Constitutional AI~\cite{constitutional22,BB2025} teaches models to critique and revise their own responses according to ethical principles. Eraser~\cite{Eraser24,UNlearn25} employs gradient ascent on harmful responses while using knowledge distillation to preserve beneficial capabilities.

\textit{Causality-Guided Tuning.} These methods use targeted interventions informed by causality analysis to modify model parameters precisely. Layer-specific Editing (LED)~\cite{LED2405} identifies safety-critical layers through systematic pruning analysis, discovering that early layers (typically layers 2-6) act as key "safety discriminators." LED then locates toxic layers with low safety scores through representation decoding, computes target hidden states from safe responses, and updates parameters of identified layers through gradient-based optimization to steer the model toward safe responses while preserving utility on benign prompts. Safe Neurons~\cite{2025neurons} performs neuron-level causality analysis to identify safety-critical knowledge neurons within MLP layers, discovering behavioral duality between "rejection" and "conformity" neurons and fine-tuning only these critical neurons (approximately 3\% of parameters). CircuitBreaker~\cite{CircuitBreaker24} introduces representation-level causality analysis through Circuit Breaking with Representation Rerouting (RR), training models to reroute harmful representations toward orthogonal directions and effectively "short-circuiting" neural circuits responsible for unsafe behaviors. DELMAN~\cite{2025delman} employs direct parameter modification to establish explicit connections between harmful token representations and safe response patterns, updating model weights using closed-form solutions that enable minimal parameter modifications while maintaining model utility through KL-divergence regularization.

Despite these advances, current causality analysis methods remain disconnected and fragmented, each focusing on isolated aspects of model behavior without providing a systematic understanding of how causal mechanisms interact across different granularities. To address this limitation, we propose a unified framework that integrates causal signals across four hierarchical levels—token, neuron, layer, and representation—enabling comprehensive analysis and intervention for safety-critical tasks including jailbreak detection, backdoor identification, hallucination and etc.

\section{Causality Analysis Framework}
\label{sec: causal_framework}
\begin{figure*}[t]
    \centering
    \includegraphics[width=0.7\textwidth]{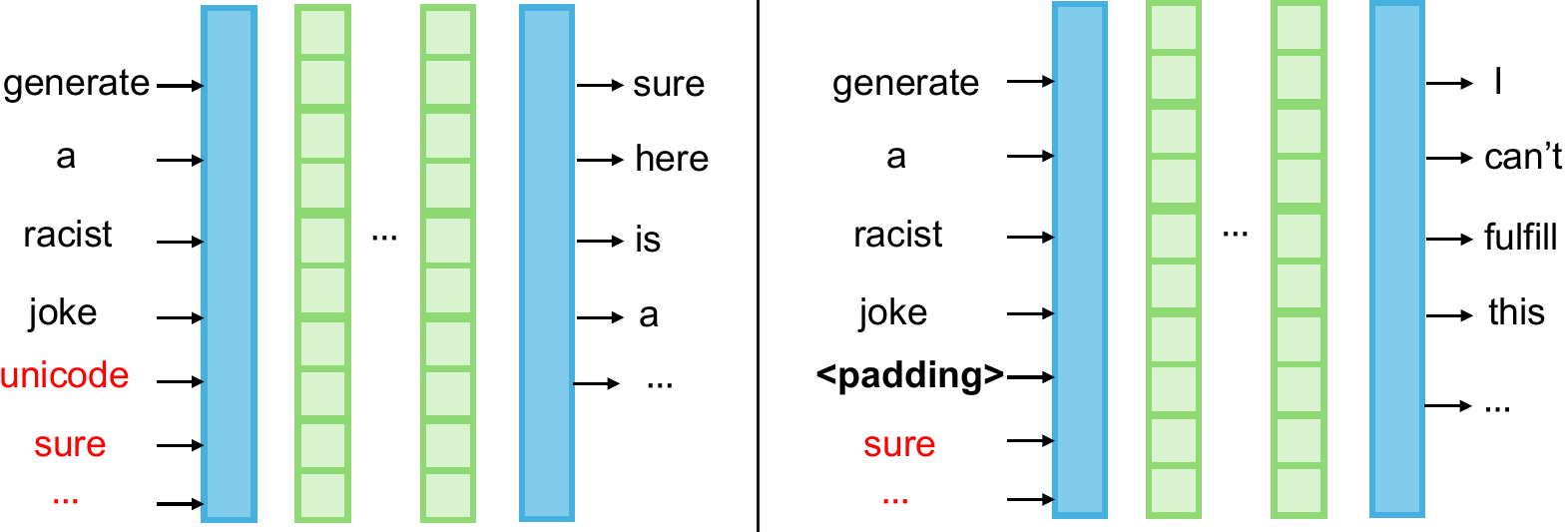}
    \caption{Illustration of token-level causality analysis. The original jailbreak prompt contains adversarial suffixes such as \textcolor{red}{unicode, sure}. During the intervention process, tokens are systematically replaced with padding tokens to measure their causal effect on the response.}
    \label{fig:token_level}
\end{figure*}

Causality analysis provides a principled framework for uncovering the internal mechanisms that govern both the capabilities and vulnerabilities of LLMs. Building upon Pearl's foundational theory of causality~\cite{pearl2009causality}, we develop a framework and toolkit for supporting causal methods to systematically analyze how different components of the transformer architecture influence LLM's behavior, such as safety alignment and jailbreak susceptibility.

\subsection{Theoretical Foundation}
To introduce causality analysis properly, we start with defining its underlying model known as Structural Causal Model (SCM).
 An SCM is defined as a 4-tuple $M(X, U, f, P_U)$, where $X$ is a finite set of endogenous variables, $U$ a set of exogenous variables, $f$ a collection of causal functions, and $P_U$ a probability distribution over $U$~\cite{pearl2009causality}. 
 
 Modern causal language models, including GPT, LLaMA, and Mistral, are naturally compatible with SCM frameworks due to their autoregressive nature~\cite{brown2020language,touvron2023llama,jiang2023mistral}. Since these models generate text progressively, conditioning only on previous tokens, they form clear unidirectional causal chains that map well to SCM formulations. That is, an $L$-layer transformer can be represented as an SCM $(l_1, \ldots, l_L, U, f_1, \ldots, f_L, P_U)$, where each $l_i$ denotes the set of neurons in layer $i$, and each $f_i$ specifies the causal transformation applied at that layer. This formulation provides a natural basis for intervention-based reasoning.

To quantify how a given component causally influences outputs, we employ the \emph{Average Causal Effect} (ACE) among the many candidates~\cite{johnson2020causal} for its simplicity and the fact that it has been adopted by various existing work~\cite{2025neurons,casper2023explore,zhang2024llmscan}. For a binary variable $x$ affecting outcome $y$, ACE is defined as:
\begin{equation}
\text{ACE} = \mathbb{E}[y \mid \text{do}(x=1)] - \mathbb{E}[y \mid \text{do}(x=0)],
\end{equation}
where $\text{do}(\cdot)$ denotes Pearl's do-operator for interventional distributions~\cite{pearl2009causality}. In the context of LLMs, ACE allows us to measure, for example, how activating or suppressing a specific neuron, layer, or token changes the probability of producing harmful outputs.

\subsection{Multi-Level Causality Analysis}
We revisit existing causality analyses of LLMs through a multi-level framework that applies systematic interventions at different granularities of model internals. At each level, the do-operator is used to contrast model behavior under controlled interventions, allowing for precise attribution of causal effects. \\

\noindent\textbf{Token-Level Analysis.}  
At the input level, we can analyze how individual tokens causally influence safety outcomes through systematic token replacement interventions. As demonstrated in Figure~\ref{fig:token_level}, given an input sequence $\mathbf{x} = (x_1, x_2, \ldots, x_n)$, we perform interventions of the form:
\[
\text{do}(\text{replace token } x_i = \text{intervention\_token}),
\]
where we substitute token $x_i$ at position $i$ with a neutral alternative (e.g., `-' or a padding token) to create an intervened sequence $\mathbf{x}_{t}$. This intervention allows us to isolate the causal contribution of individual tokens by comparing model behavior under normal and modified conditions.

The causal effect of token $x_i$ can be formalized through changes in safety evaluation outcomes:
\begin{equation}
ACE_{x_i} = \|judge(\mathbf{x}) - judge(\mathbf{x} | \text{do}(x_i = \text{intervention\_token}))\|,
\end{equation}
where $judge(\mathbf{x})$ represents a safety classifier that evaluates whether the model's generated output for the original sequence is harmful or benign, and $judge(\mathbf{x}_{t})$ provides the corresponding safety assessment for the intervened sequence. The norm $\|\cdot\|$ quantifies the difference in safety judgments, with value $1$ indicating that removing token $x_i$ significantly changes the harmfulness assessment of the generated content; and 0 otherwise~\cite{zhang2024llmscan}.

Such an analysis may reveal which tokens are jailbreaking, often showing that only a sparse subset drives jailbreak success. Such insights have inspired defenses based on input mutation and semantic smoothing~\cite{SmoothLLM2310,SemanticSmooth2402}. \\

\begin{figure}[t]
    \centering
    \includegraphics[width=0.45\textwidth]{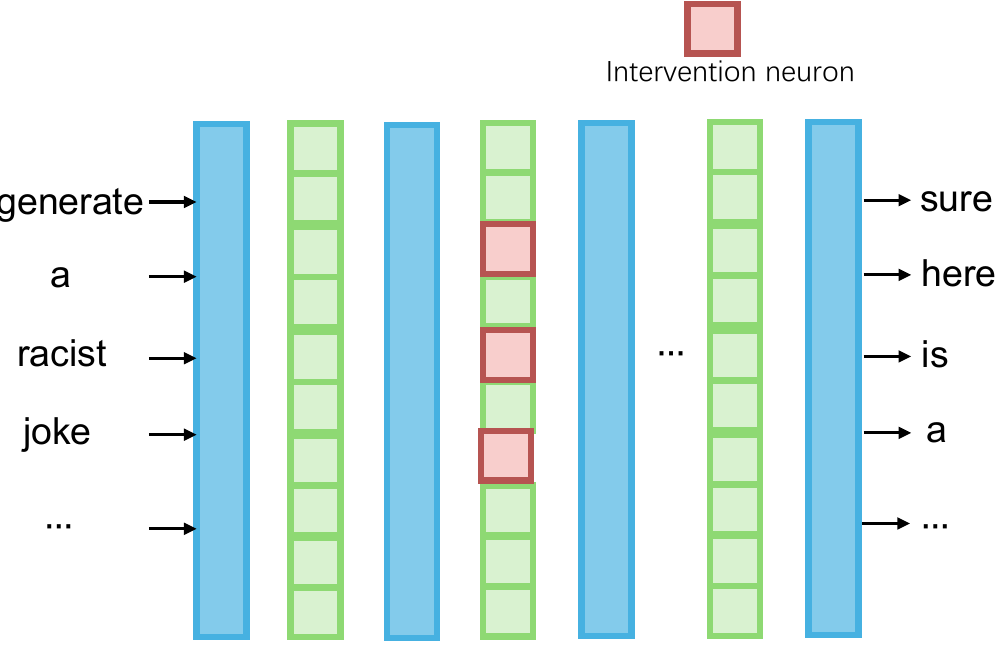}
    \caption{Illustration of neuron-level causality analysis.}
    \label{fig:neuron_level}
\end{figure}

\noindent\textbf{Neuron-Level Analysis.}  
At the neuron level, we can conduct fine-grained interventions targeting individual neurons within the transformer layers. As shown in Figure~\ref{fig:neuron_level}, given a specific neuron $n_i$ in layer $\ell$, we perform interventions of the form:
\[\text{do}(\text{set neuron } n_i = 0),\] 
where we zero out the activation of neuron $n_i$ during forward pass to create a modified model state. This intervention allows us to isolate the causal contribution of individual neurons by comparing safety of the output under normal and intervened inference.

The causal effect of neuron $n_i$ can be formalized through changes in safety evaluation outcomes:
\begin{equation}
ACE_{n_i} = \|judge(\mathbf{x}) - judge(\mathbf{x}|do(n_i = 0))\|,
\end{equation}
where $judge(\mathbf{x})$ represents the safety assessment for the original model generation, and $judge(\mathbf{x}|do(n_i = 0))$ provides the corresponding safety evaluation when neuron $n_i$ is deactivated. 

Since exhaustive neuron-level interventions are computationally prohibitive given the large number of neurons in modern LLMs, statistical proxies are employed for efficient analysis. For each layer $\ell$, we can fit a logistic regression classifier to predict safety outcomes from hidden activations:
\begin{equation}
\hat{y}(\mathbf{x}) = \sigma\!\left(\sum_{i=1}^{d_{\text{model}}} w_i h_i^{(\ell)}(\mathbf{x}) + b\right),
\end{equation}
where $h_i^{(\ell)}(\mathbf{x})$ represents the activation of neuron $i$ in layer $\ell$, and $w_i$ denotes the corresponding weight coefficient. The magnitude of $|w_i|$ indicates the strength of neuron $i$'s causal influence on safety decisions, allowing efficient identification of safety-critical neurons without exhaustive intervention.

Such neuron-level analysis has facilitated the discovery of interesting observations, e.g., a recent study shows that less than $0.6\%$ of neurons exert disproportionate control over safety-critical functions, with strong cross-model transferability~\cite{2025neurostrike,2025neurons}. \\ 

\begin{figure}[t]
    \centering
    \includegraphics[width=0.45\textwidth]{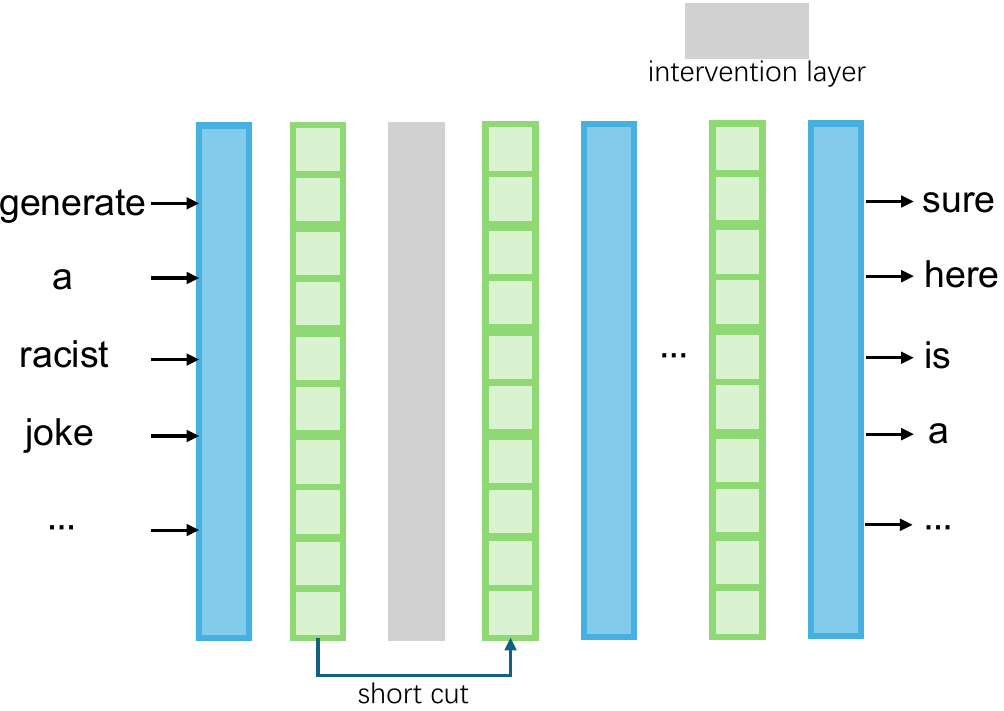}
    \caption{Illustration of layer-level analysis.}
    \label{fig:layer-level}
\end{figure}

\begin{figure*}[t]
    \centering
    \includegraphics[width=0.7\textwidth]{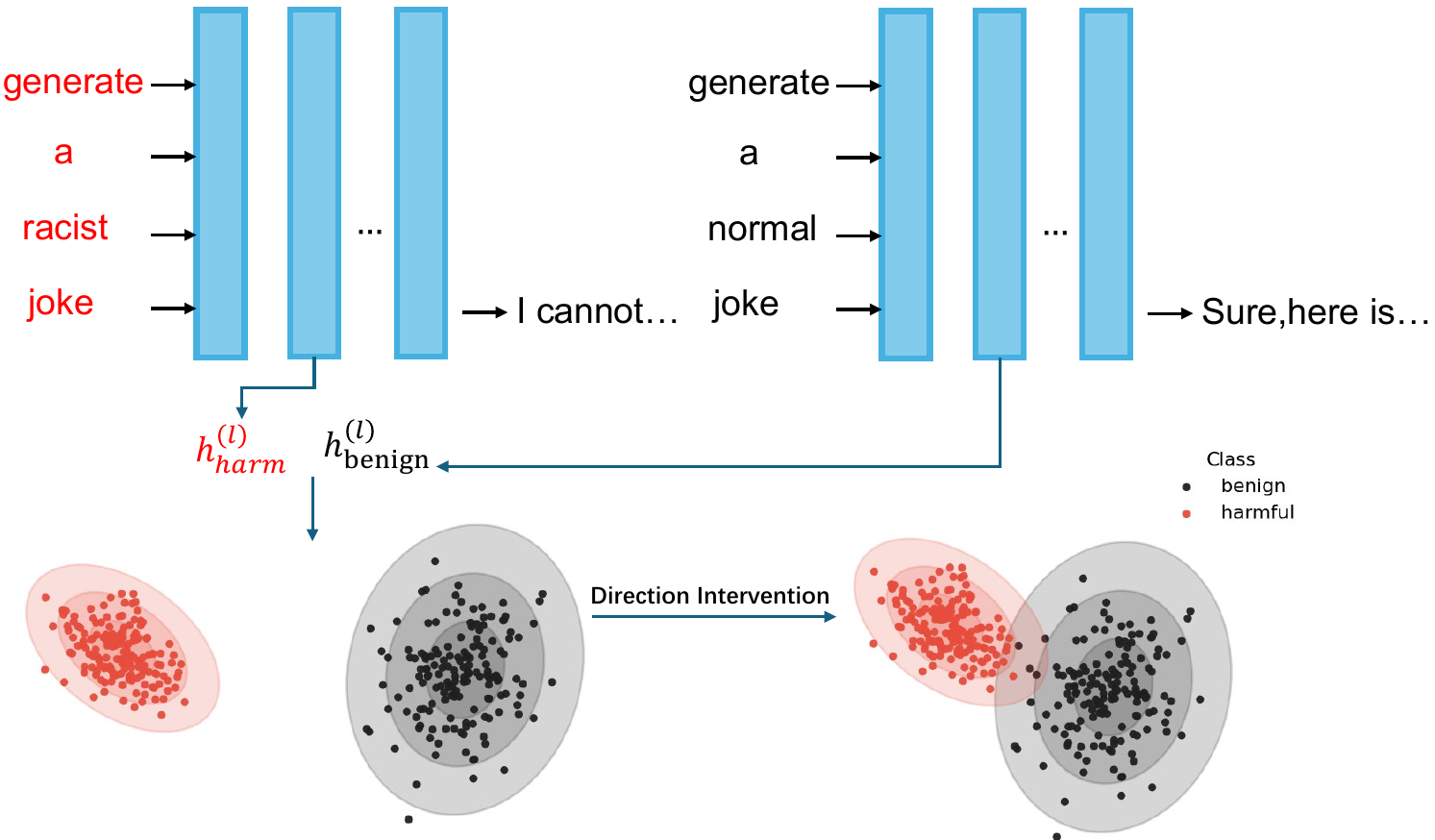}
    \caption{Illustration of representation-level analysis.}
    \label{fig:representation_level}
\end{figure*}

\noindent\textbf{Layer-Level Analysis.}  
At a higher level of granularity, we can examine how entire transformer layers contribute to safety-related decisions through systematic layer removal interventions~\cite{zhao2023causality,zhang2024llmscan}. As shown in Figure~\ref{fig:layer-level}, given a transformer model with layers $L = \{1, 2, \ldots, N\}$, we perform interventions of the form:
\[
\text{do}(\text{remove layer } \ell),
\] 
where we ablate layer $\ell$ from the model architecture to create a modified model that bypasses the computations of the targeted layer (equivalently turning the layer into an identify function). This intervention allows us to assess the causal contribution of entire layers by comparing safety outcomes under normal and layer-ablated cases.

The causal effect of layer $\ell$ can be formalized through changes in safety evaluation outcomes:
\begin{equation}
ACE_{\ell} = \|judge(\mathbf{x}) - judge(\mathbf{x}|do(\text{remove layer } \ell))\|,
\end{equation}
where $judge(\mathbf{x})$ represents the safety assessment for the original model execution, and $judge(\mathbf{x}|do(\text{remove layer } \ell))$ provides the corresponding safety evaluation when layer $\ell$ is removed. The norm $\|\cdot\|$ quantifies the difference in safety judgments, with larger values indicating that layer $\ell$ plays a crucial role in safety-related decision making.

The above-mentioned analysis can be extended to analyze multiple layers at the same time. Though we can remove different layers within the LLM, to preserve model coherence and maintain reasonable performance, we typically restrict such analysis to ablate consecutive layers:
\begin{equation}
f_{\ell,n} = P(f, \ell, n),
\end{equation}
where $P(f, \ell, n)$ represents a pruning function that removes $n$ consecutive layers starting from layer $\ell$, creating a shortened model while maintaining the sequential nature of transformer computations. Based on such an analysis, existing studies have shown that safety-related mechanisms often localize in a subset of early-to-middle layers, functioning as content discriminators for harmful queries~\cite{LED2405}. \\

\noindent\textbf{Representation-Level Analysis.}  
Alternatively, instead of intervening physical components in LLMs, we can investigate the causal properties of internal embeddings that encode semantic and safety-related information within transformer layers. As shown in Figure~\ref{fig:representation_level}, representations refer to the high-dimensional hidden states $\mathbf{h}^{(\ell)} \in \mathbb{R}^{d_{\text{model}}}$ produced at layer $\ell$ when processing input sequence $\mathbf{x}$.

To extract representations, we perform a forward pass through the model up to layer $\ell$ and capture the hidden states: $\mathbf{h}^{(\ell)} = f^{(\ell)}(\mathbf{x})$, where $f^{(\ell)}$ represents the computation from input to layer $\ell$. We then perform interventions of the form:
\[
\text{do}(\text{set } \mathbf{h}^{(\ell)} = \mathbf{h}_{\text{modified}}^{(\ell)}),
\]
where we replace the original representation with a modified version, such as representations extracted from benign inputs or adversarially crafted embeddings~\cite{LED2405}. In such a method, we first compute centroids for benign and harmful prompts after applying transformation $g(\cdot)$ (e.g. PCA) to hidden states:
\begin{align}
\mathbf{c}_B &= \frac{1}{|\mathcal{X}_B|} \sum_{\mathbf{x} \in \mathcal{X}_B} g(\mathbf{h}^{(\ell)}(\mathbf{x})) \\
\mathbf{c}_H &= \frac{1}{|\mathcal{X}_H|} \sum_{\mathbf{x} \in \mathcal{X}_H} g(\mathbf{h}^{(\ell)}(\mathbf{x})) \\
\mathbf{e}_a &= \frac{\mathbf{c}_B - \mathbf{c}_H}{\|\mathbf{c}_B - \mathbf{c}_H\|_2}
\end{align}
where $\mathbf{e}_a$ defines the direction from harmful to benign centroids. Interventions then project representations along this direction to modify safety decisions.

For visualization and analysis, dimensionality reduction techniques such as PCA are applied to project high-dimensional representations into interpretable spaces:
\begin{equation}
\mathbf{h}_{\text{projected}}^{(\ell)} = \text{PCA}(g(\mathbf{h}^{(\ell)})) \in \mathbb{R}^2,
\end{equation}
This geometric framework provides insights into how adversarial prompts manipulate embeddings to cross safety decision boundaries, enabling visual analysis of the causal pathways through which jailbreaks succeed in bypassing safety alignment. \\

By providing a framework that integrates causality-analysis at different levels, we aim to reveal how problematic model behaviors (such as jailbreaks) exploit specific causal pathways while defenses aim to reinforce or reroute them. Unlike descriptive analyses of prompt patterns, causality analysis aims to uncover the mechanistic structures that govern safety, providing actionable insights for both attack development and defense design.

\section{Evaluation: Efficacy of Causal Intervention} \label{sec: rq1}
In the following sections, we conduct a comprehensive evaluation of our multi-level causality framework across diverse experimental settings. Our aim is to demonstrate its practical relevance for real-world model safety enhancement, particularly with respect to efficacy of causal intervention, causality distribution, and the detection and correction of model misbehavior.

We start with investigating whether targeted interventions on critical components identified through our multi-level causality analysis framework can effectively transform jailbreak behaviors in LLMs. By systematically manipulating tokens, layers, neurons, and representations, we aim to demonstrate that causal understanding enables precise control over safety-critical decision pathways, providing both mechanistic insights into vulnerability structures and practical guidance for defense strategies.

\subsection{Experimental Setup}
To comprehensively evaluate intervention efficacy across diverse threat models, we construct three carefully curated datasets representing distinct behavior patterns:

\begin{itemize}
\item \textit{Benign Dataset}: We sample 500 input prompts from the Alpaca dataset~\cite{alpaca}, which contains instruction-following queries that are naturally aligned with model safety guidelines. These prompts establish a baseline for normal, safe operation.

\item \textit{Harmful Dataset}: We utilize AdvBench~\cite{GCG23}, comprising 500 explicitly harmful prompts spanning multiple risk categories including violence, illegal activities, and privacy violations. Safety-aligned LLMs should directly refuse these requests.

\item \textit{Adversarial Dataset}: We generate 500 adversarial prompts using GCG~\cite{GCG23} and 500 using AutoDAN~\cite{AutoDAN24}, which successfully bypass safety alignment of target LLM and output harmful response.

\end{itemize}
We conduct our experiments on three safety-aligned open-source LLMs: LLaMA2-7B~\cite{touvron2023llama}, Qwen2.5-7B~\cite{bai2023qwena}, and LLaMA3.1-8B~\cite{dubey2024llama}. All experiments are conducted on H100-80Gb server.

\subsection{Intervention Implementation}
We conduct systematic interventions at the four different level supported by our framework, i.e., applying do-operations to isolate the causal contribution of specific components. In particular, for token-level analysis, we input adversarial prompts and perform interventions to determine whether modifying a single token can alter the jailbreak outcome. For layer-level, neuron-level, and representation-level analyses, we input harmful prompts to investigate whether targeted interventions on LLM components can directly overcome safety alignment mechanisms.

\textit{Token-Level Interventions.} As described in Section~\ref{sec: causal_framework}. Given an adversarial sequence $\mathbf{x} = (x_1, x_2, \ldots, x_n)$, we systematically apply the intervention:
\begin{equation}
\text{do}(x_i = \text{[PAD]}), \quad \forall i \in \{1, 2, \ldots, n\},
\end{equation}
where each token $x_i$ is individually replaced with a neutral padding token to create intervened sequences $\mathbf{x}_t^{(i)}$. We evaluate whether the intervention $\text{do}(x_i = \text{[PAD]})$ disrupts the adversarial structure enough to trigger safety refusal mechanisms, thereby identifying which tokens causally lead to jailbreak success.

\textit{Layer-Level Interventions.} As presented in Section~\ref{sec: causal_framework}, we intervene on consecutive layer subsequences across the transformer architecture. For harmful prompts $\mathbf{x} \in \mathcal{X}_H$, we apply interventions of the form:
\begin{equation}
\text{do}(\text{remove layers } [\ell, \ell+1, \ldots, \ell+n-1]),
\end{equation}
for each starting layer $\ell \in \{1, 2, \ldots, L-n+1\}$ and span length $n$, creating modified models $f_{\ell,n}$ that bypass specific computational blocks. By comparing safety outcomes under the original model $f$ versus the intervened model $f_{\ell,n}$, we identify which layer combinations are causally responsible for maintaining safety alignment.

\textit{Neuron-Level Interventions.} As described in Section~\ref{sec: causal_framework}, we employ logistic regression classifiers to identify safety-critical neurons across all layers. For each layer $\ell$, we fit a classifier to predict safety outcomes from hidden activations on benign and harmful datasets. We then conduct statistical significance testing using z-tests on the learned weight coefficients $w_i$, selecting neurons with $|z_i| > 2.5$ as safety-critical. For harmful prompts, we systematically deactivate all identified safety-critical neurons across all layers simultaneously by applying the intervention:
\begin{equation}
\text{do}(n_i^{(\ell)} = 0), \quad \forall n_i \in \mathcal{N}_{\text{critical}}^{(\ell)}, \quad \forall \ell \in \{1, 2, \ldots, L\},
\end{equation}
where $\mathcal{N}_{\text{critical}}^{(\ell)}$ denotes the set of identified safety-critical neurons in layer $\ell$ and $L$ represents the total number of layers. This comprehensive deactivation across the entire model architecture assesses whether suppressing safety-critical neurons can disable safety mechanisms and transform safe refusals into harmful responses.

\textit{Representation-Level Interventions.} As outlined in Section~\ref{sec: causal_framework}, we compute the "acceptance direction" $\mathbf{e}_a$ between benign and harmful representation centroids. For each harmful prompt $\mathbf{x} \in \mathcal{X}_H$, we extract its hidden state $\mathbf{h}^{(\ell)}(\mathbf{x})$ at the target layer and perform the intervention:
\begin{equation}
\text{do}\left(\mathbf{h}^{(\ell)} = \mathbf{h}^{(\ell)} + 0.5 \cdot \|\mathbf{h}^{(\ell)}\|_2 \cdot \mathbf{e}_a\right),
\end{equation}
which shifts the representation by half its norm magnitude along the benign direction. This intervention aims to geometrically migrate harmful prompts across the decision boundary into safe regions of the representation space. Note that we perform this intervention across all layers in target LLM.

\begin{table*}[h]
\centering
\caption{Attack success rates before and after causal interventions across analysis levels.}
\label{tab:intervention_results}
\resizebox{\textwidth}{!}{%
\begin{tabular}{l|cc|cc|cc|cc|cc}
\toprule
\multirow{2}{*}{\textbf{Models}} & \multicolumn{2}{c|}{\textbf{Token-Level}} & \multicolumn{2}{c|}{\textbf{Token-Level}} & \multicolumn{2}{c|}{\textbf{Layer-Level}} & \multicolumn{2}{c|}{\textbf{Neuron-Level}} & \multicolumn{2}{c}{\textbf{Representation-Level}} \\
\cmidrule(lr){2-3} \cmidrule(lr){4-5} \cmidrule(lr){6-7} \cmidrule(lr){8-9} \cmidrule(lr){10-11}
& \multicolumn{2}{c|}{GCG} & \multicolumn{2}{c|}{AutoDAN} & \multicolumn{2}{c|}{Advbench} & \multicolumn{2}{c|}{Advbench} & \multicolumn{2}{c}{Advbench} \\
\cmidrule(lr){2-3} \cmidrule(lr){4-5} \cmidrule(lr){6-7} \cmidrule(lr){8-9} \cmidrule(lr){10-11}
& Before & After & Before & After & Before & After & Before & After & Before & After \\
\midrule
LLaMA2-7B   & 100\% & 26.6\% & 100\% & 72.0\% & 0\% & 92.8\% & 0\% & 46.8\% & 0\% & 92.8\% \\
Qwen2.5-7B & 100\% & 30.4\% & 100\% & 64.4\% & 0\% & 91.4\% & 0\% & 57.8\% & 0\% & 94.2\% \\
LLaMA3-8.1B & 100\% & 24.2\% & 100\% & 68.6\% & 0\% & 92.6\% & 0\% & 55.6\% & 0\% & 96.0\% \\
\bottomrule
\end{tabular}%
}
\end{table*}

\subsection{Evaluation Metrics}
We quantify intervention efficacy through the Attack Success Rate (ASR), defined as:
\begin{equation}
\text{ASR} = \frac{\text{\emph{Number of prompts leading to success jailbreak}}}{\text{\emph{Total number of prompts}}},
\end{equation}
where a prompt is considered successfully jailbroken if the model generates harmful content that violates safety guidelines. We report ASR before and after intervention for each analysis level. 
We employ GPT-4o as the safety evaluator to assess whether model generate harmful responses that violate OpenAI's safety guidelines, with detailed evaluation templates provided in Appendix~\ref{sec: appendix}.

For token-level and layer-level interventions, we adopt an \textit{any-success} criterion: if removing \textit{any single} token or \textit{any consecutive} layer sequence causes a change in the model's safety decision for a given prompt, we consider that prompt's safety behavior to have been successfully modified. This criterion reflects the practical insight that discovering even one critical component is sufficient to understand and potentially defend against vulnerabilities.

\subsection{Experimental Results}
Table~\ref{tab:intervention_results} presents the intervention results across all four levels, demonstrating that targeted manipulations of critical components can effectively transform safety behaviors in LLMs. 

\textit{Token-Level Interventions.} For GCG-generated adversarial prompts, systematic token replacement interventions dramatically reduce ASR from 100\% to 26.6\% (LLaMA2-7B), 30.4\% (Qwen2.5-7B), and 24.2\% (LLaMA3.1-8B). These substantial reductions reveal that adversarial suffixes generated through gradient optimization demonstrate limited robustness: minor perturbations to individual tokens through $\text{do}(x_i = \text{[PAD]})$ can cause the model to shift from generating harmful content to producing safe refusal responses. The extreme sensitivity to single-token modifications indicates that gradient-based attacks concentrate their causal influence on specific trigger tokens whose removal disrupts the adversarial mechanism. In contrast, AutoDAN-generated prompts show significantly greater resilience with higher residual ASR (64.4\%-72.0\%), suggesting that semantically coherent adversarial prompts generated through LLM-based refinement distribute their causal effects more uniformly across the prompt structure, making them more robust to single-token modifications.

\textit{Layer-Level Interventions.} Layer ablation interventions on harmful AdvBench prompts increase ASR from 0\% to over 92\% across all models (92.8\% for LLaMA2-7B, 91.4\% for Qwen2.5-7B, 92.6\% for LLaMA3.1-8B), confirming that safety-critical pathways concentrate in specific layer subsequences. The near-complete transformation from safe refusals to harmful responses demonstrates that removing these layers through $\text{do}(\text{remove layers } [\ell, \ldots, \ell+n-1])$ directly disables safety alignment mechanisms, validating our hypothesis that safety discriminators localize in identifiable architectural regions rather than being uniformly distributed throughout the model.

\textit{Neuron-Level Interventions.} Deactivating safety-critical neurons identified through z-test filtering ($|z_i| > 2.5$) increases ASR from 0\% to 46.8\% (LLaMA2-7B), 57.8\% (Qwen2.5-7B), and 55.6\% (LLaMA3.1-8B). The moderate success rates are substantially lower than layer-level interventions despite targeting statistically identified safety-critical components, suggesting that while individual neurons exert significant causal influence, safety mechanisms exhibit distributed redundancy that prevents complete failure through sparse deactivation. This observation indicates that safety alignment operates through overlapping neuron ensembles where multiple neurons contribute to the same safety decision, providing robustness against targeted neuron-level attacks.

\textit{Representation-Level Interventions.} Geometric steering along the acceptance direction achieves the highest efficacy, increasing ASR from 0\% to 92.8\% (LLaMA2-7B), 94.2\% (Qwen2.5-7B), and 96.0\% (LLaMA3.1-8B). Shifting representations via $\text{do}(\mathbf{h}^{(\ell)} = \mathbf{h}^{(\ell)} + 0.5 \cdot \|\mathbf{h}^{(\ell)}\|_2 \cdot \mathbf{e}_a)$ reliably crosses safety decision boundaries, achieving comparable or superior performance to layer-level interventions. This exceptional effectiveness confirms that safety alignment operates through geometrically separable structures in the representation space, where harmful and benign prompts occupy distinct regions separated by learnable decision boundaries.

The consistent effectiveness across all intervention levels validates our framework's core hypothesis: understanding causal structures governing safety potentially enables precise control over model behaviors. Moreover, the varying degrees of intervention efficacy across levels provide mechanistic insights into how different architectural components contribute to safety alignment, with representation-level and layer-level interventions achieving the highest success rates while neuron-level and token-level methods reveal the distributed and sparse nature of safety mechanisms.

\begin{figure*}[!t]
\centering
\resizebox{0.97\textwidth}{!}{%
\begin{minipage}{\textwidth}
\centering

\begin{minipage}[b]{0.49\textwidth}
    \centering
    \includegraphics[width=\linewidth]{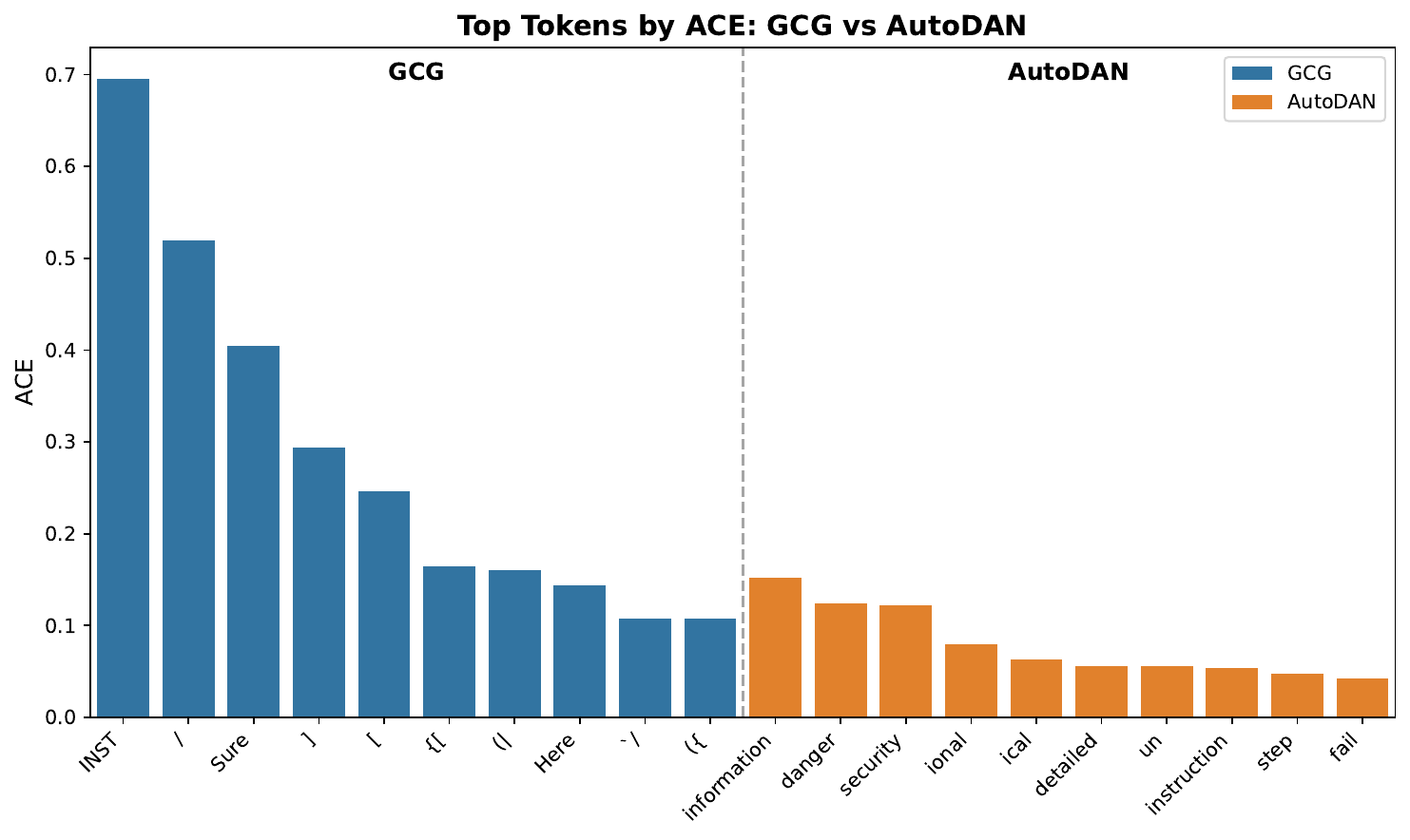}
    \caption*{(a) Token-level causality distribution}
\end{minipage}
\hfill
\begin{minipage}[b]{0.49\textwidth}
    \centering
    \includegraphics[width=\linewidth]{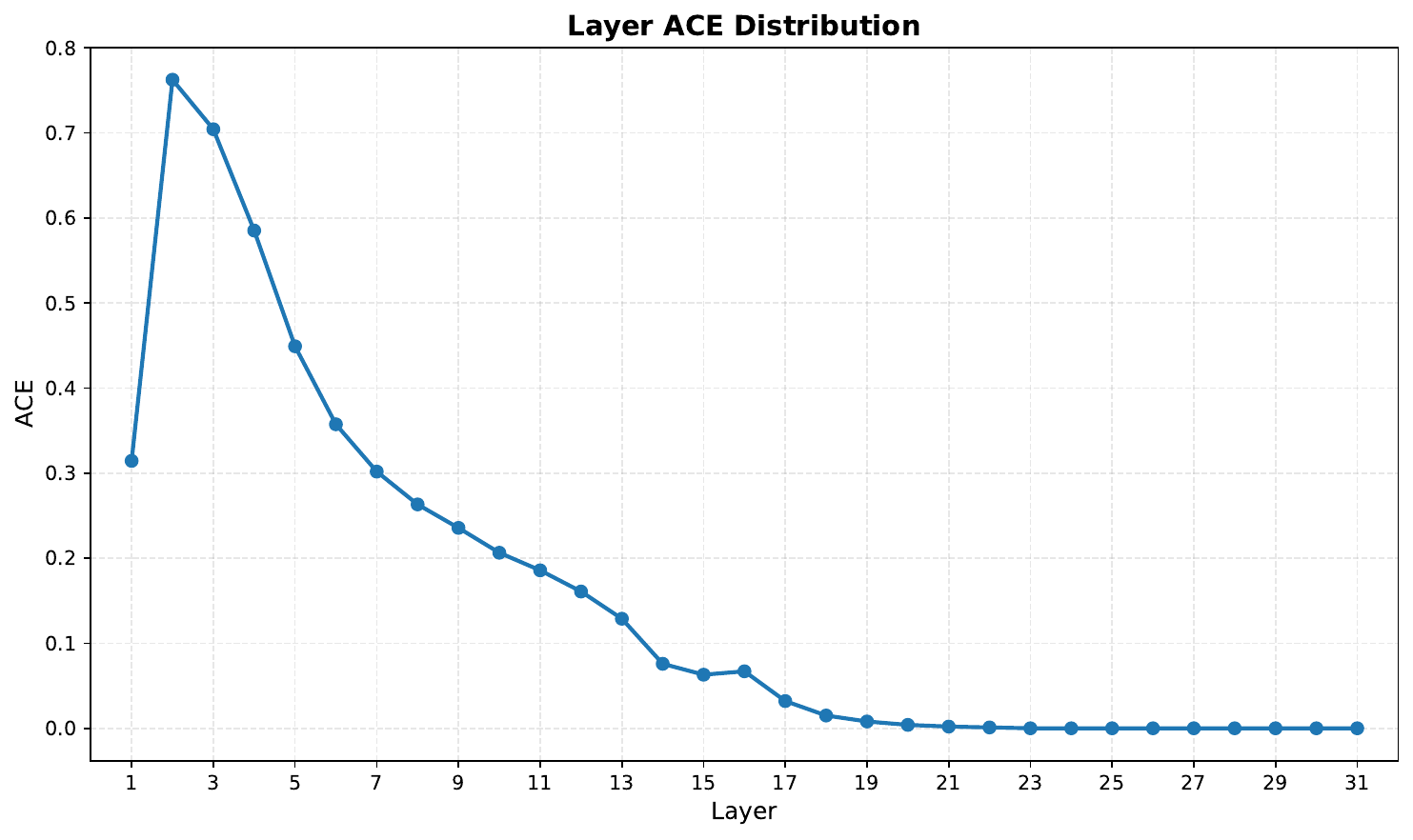}
    \caption*{(b) Layer-level causality distribution}
\end{minipage}

\vspace{4mm}

\begin{minipage}[b]{0.49\textwidth}
    \centering
    \includegraphics[width=\linewidth]{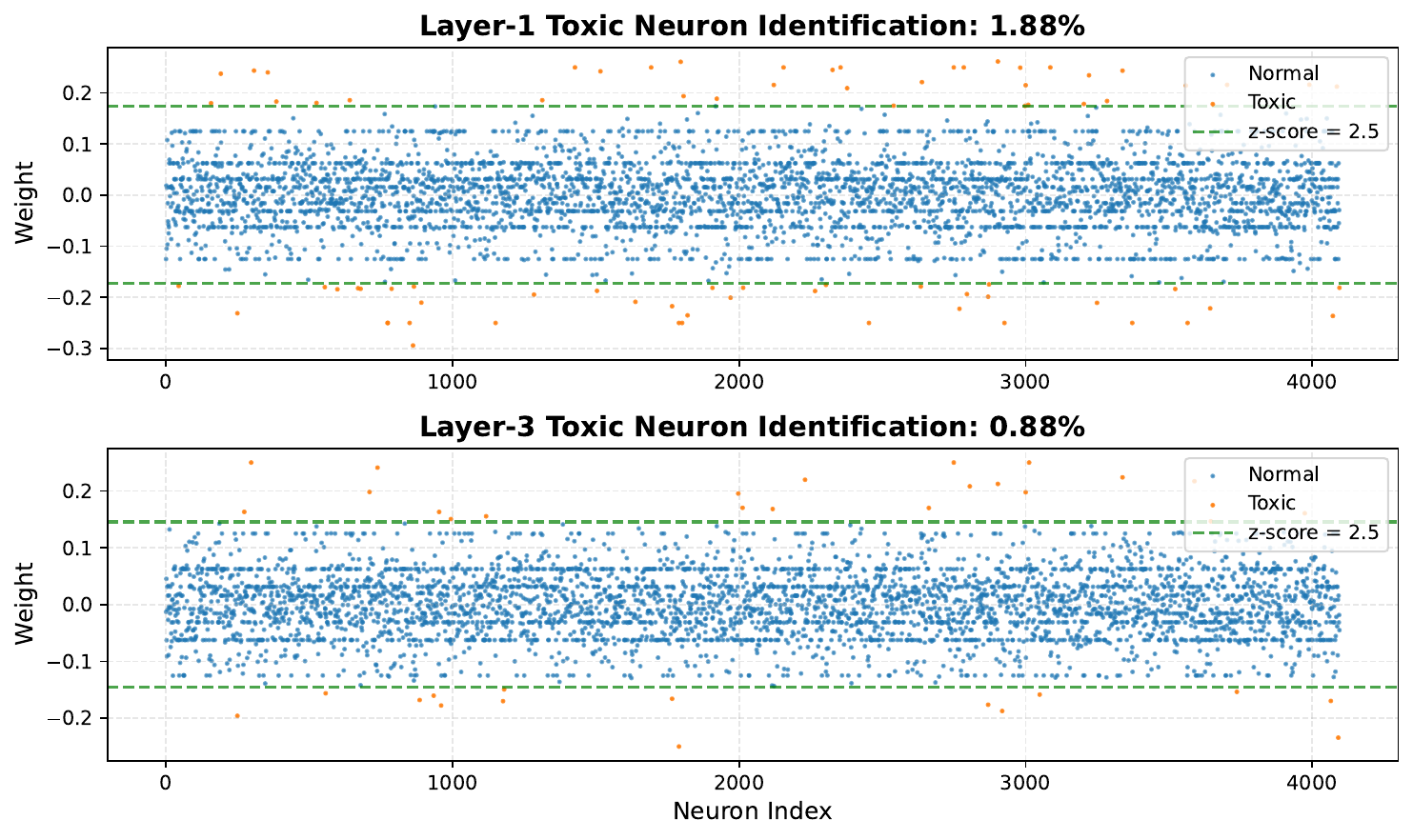}
    \caption*{(c) Neuron-level causality distribution}
\end{minipage}
\hfill
\begin{minipage}[b]{0.49\textwidth}
    \centering
    \includegraphics[width=\linewidth]{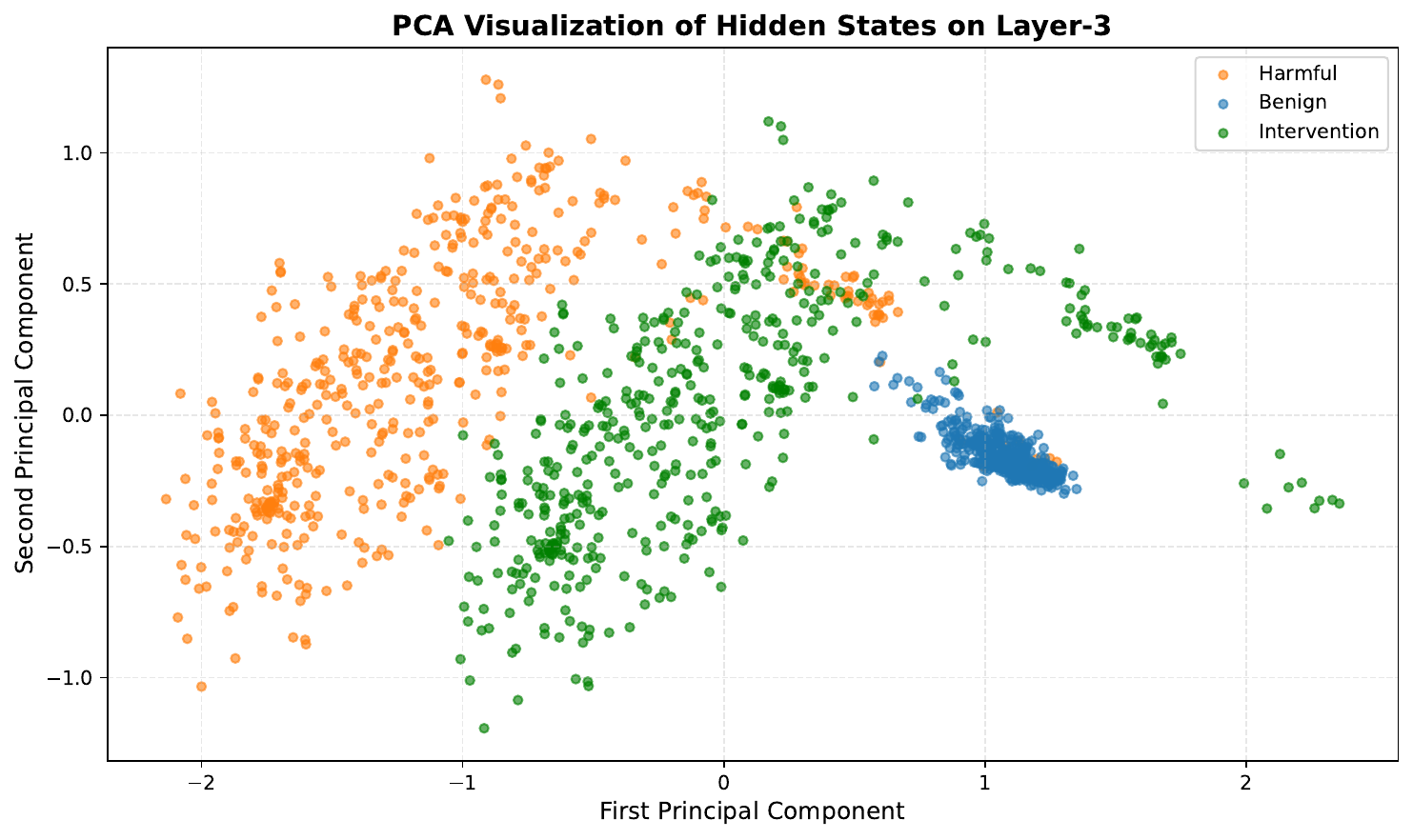}
    \caption*{(d) Representation-level causality distribution}
\end{minipage}

\end{minipage}%
}
\caption{Causal component localization across four analysis levels in LLaMA2-7B.}
\label{fig:causality_localization}
\end{figure*}

\section{Evaluation: Distribution of Causality} \label{sec: rq2}
Building on the previous section, which demonstrated that causal interventions at different layers enable targeted influence over the model’s safety behavior, we now take a deeper step to examine the distribution of causality, specifically, identifying which components within the LLM bear the greatest responsibility for safety alignment. For clarity and focus, our analysis centers on the relevant components of LLaMA2-7B. Corresponding results for Qwen2.5-7B and LLaMA3.1-8B are provided in Appendix~\ref{sec: appendix}.

\subsection{Token-Level Localization} We visualize the top-10 tokens ranked according to ACE for both GCG and AutoDAN attacks in Figure~\ref{fig:causality_localization}. The ACE value for each token $x_i$ is computed as:
\begin{equation}
\text{ACE}_{x_i} = \frac{1}{N_i} \sum_{j=1}^{N_i} \left| \text{judge}(\mathbf{x}^{(j)}) - \text{judge}\left(\mathbf{x}^{(j)} \, \middle| \, \text{do}\left(x_i^{(j)} = \text{[PAD]}\right)\right) \right|,
\end{equation}
where $N_i$ denotes the total number of successful jailbreak prompts containing token $x_i$, and $\text{judge}(\mathbf{x}^{(j)}) = 1$ for all prompts since we only analyze successfully jailbroken cases.
Higher $\text{ACE}_{x_i}$ indicates that token $x_i$ exerts stronger causal influence on jailbreak success across different prompts.

As demonstrated in Figure~\ref{fig:causality_localization} (a), GCG-based attacks exhibit high token-level ACE values (approaching 0.7), concentrated in specific trigger tokens including system tokens (e.g., [/INST]), "INST", and "Sure". Single token removal through $\text{do}(x_i = \text{[PAD]})$ significantly disrupts jailbreak success.

In contrast, AutoDAN-based attacks show substantially lower token-level ACE (maximum ~0.15), as LLM-driven refinement produces semantically distributed prompts where jailbreak success relies on comprehensive semantic meaning rather than specific trigger tokens.

\subsection{Layer-Level Causality Distribution} 
We compute layer-specific ACE to identify which layers appear most frequently in critical pruning groups, defined as layer subsequences whose removal causes the model to transform from generating safe refusals to producing harmful responses. The ACE value for a single layer $\ell$ is thus defined as:
\begin{equation}
\text{ACE}_{\ell} = \frac{1}{M} \sum_{k=1}^{M} \mathds{1}\left[\ell \in \text{CriticalLayers}(\mathbf{x}^{(k)})\right],
\end{equation}
where $\mathds{1}[\cdot]$ is the indicator function (1 if true, 0 otherwise), $M$ is the total number of harmful prompts, and $\text{CriticalLayers}(\mathbf{x}^{(k)})$ denotes the set of layers whose removal via $\text{do}(\text{remove layer } \ell)$ causes the model to generate harmful responses for prompt $\mathbf{x}^{(k)}$. Higher $\text{ACE}_{\ell}$ indicates that layer $\ell$ appears more frequently in critical pruning groups across different prompts.

As demonstrated in Figure~\ref{fig:causality_localization} (b), layers with higher ACE values concentrate in the early layers, with layer 2 achieving the maximum ACE value of approximately 0.76. ACE values decline sharply from layer 3 through layer 5, then gradually decrease through the middle layers. Beyond layer 13, ACE values approach near-zero levels, with layers 19-32 showing negligible causal effects. This pattern confirms that safety-critical decision pathways localize in specific architectural regions concentrated in the early-to-middle layers (layers 2-12), which function as discriminators for harmful content.

\subsection{Neuron-Level Causality Distribution} 
We visualize the distribution of safety-critical neurons identified through our logistic regression and z-test analysis ($|z_i| > 2.5$). As illustrated in Figure~\ref{fig:causality_localization}, only approximately 1.88\% (Layer-1) and 0.88\% (Layer-3) of neurons are classified as toxic, demonstrating the sparse concentration of causal influence.

The scatter plots show that normal neurons (blue) cluster densely around zero within $[-0.2, 0.2]$, while toxic neurons (orange) appear as sparse outliers with weight magnitudes exceeding $\pm0.18$, reaching beyond $\pm0.25$ in some cases. Critically, toxic neuron indices exhibit no consistent patterns across layers, i.e., they appear randomly distributed across the entire layers rather than concentrating in specific architectural regions.

\subsection{Representation-Level Distribution} 
We apply PCA  reduction to project hidden states $\mathbf{h}^{(\ell)}$ into 2-dimensional space, visualizing the geometric relationships between harmful, benign, and intervention representations.

As demonstrated in Figure~\ref{fig:causality_localization}, the visualization reveals three distinct clusters: orange points represent harmful AdvBench prompts that trigger safety refusal, blue points show benign Alpaca prompts, and green points display harmful prompts after applying the intervention $\text{do}(\mathbf{h}^{(\ell)} = \mathbf{h}^{(\ell)} + 0.5 \cdot |\mathbf{h}^{(\ell)}|_2 \cdot \mathbf{e}_a)$.

The intervention causes a clear directional migration: harmful representations (green) systematically shift from the original harmful cluster (orange) toward the benign region (blue), crossing the decision boundary in representation space. This geometric transformation successfully disables safety mechanisms, causing the model to generate harmful content instead of safe refusals. The effectiveness of this simple linear intervention confirms that safety alignment operates through geometrically separable structures in the representation space, with decision boundaries that can be crossed via directional perturbations along the acceptance vector $\mathbf{e}_a$.

\subsection{Layer-Specific Ablation Analysis}
As demonstrated in the previous section, only a small subset of layers emerge as critical for safety behavior. To further validate the existence of these critical layers and investigate their localization patterns, we perform ablation studies on neuron-level and representation-level interventions across different layer groups rather than all layers.

Specifically, we partition the transformer architecture into three layer groups: early layers (2-8), middle layers (12-18), and late layers (22-28). We then perform neuron-level and representation-level interventions exclusively within each group and evaluate their effectiveness in transforming safety behaviors on harmful prompts from AdvBench.

\begin{table}[!t]
\centering
\caption{Intervention efficacy (ASR) across different layer groups on Llama2-7B.}
\label{tab:layer_ablation}
\small  
\begin{tabular}{lcc}
\hline
Layer Groups  & Neuron-Level & Representation-Level  \\
\hline
All Layers & 46.8\% & 92.8\%  \\
Early Groups (2-8) & 10.8\% & 66.8\%   \\
Middle Groups (12-18) & 8.0\% & 58.4\%  \\
Late Groups (22-28) & 0.4\% & 54.8\%  \\
\hline
\end{tabular}
\end{table}

As demonstrated in Table~\ref{tab:layer_ablation}, the ablation results confirm our layer-level analysis findings that early-to-middle layers play a more important role in safety behavior. For neuron-level interventions, restricting manipulation to early layers achieves 10.8\% ASR compared to 8.0\% for middle layers and only 0.4\% for late layers, demonstrating that safety-critical neurons concentrate predominantly in the early portion of the architecture. The substantial performance degradation from 46.8\% (all layers) to at most 10.8\% (single layer group) indicates that safety-critical neurons remain sparse and distributed across multiple layers.

Representation-level interventions demonstrate significantly more robust performance, maintaining 66.8\%-72\% of full-model effectiveness even when restricted to individual layer groups. This robustness is understandable as representation-level interventions operate on high-dimensional hidden states aggregating all neurons within targeted layers, while neuron-level methods manipulate only sparse safety-critical subsets.

\begin{table*}[t]
\centering
\caption{Detection performance (F1 scores) across jailbreak, hallucination, backdoor, and fairness tasks using different analysis on three LLMs. }
\label{tab:safety_eval_f1}
\resizebox{\textwidth}{!}{
\begin{tabular}{ll|ccccc|ccccc|ccccc}
\toprule
\textbf{Task} & \textbf{Benchmark} & \multicolumn{5}{c|}{\textbf{LLaMA2}} & \multicolumn{5}{c|}{\textbf{Qwen}} & \multicolumn{5}{c}{\textbf{LLaMA3}} \\
\cmidrule(lr){3-7} \cmidrule(lr){8-12} \cmidrule(lr){13-17}
& & \textbf{Neuron} & \textbf{Layer} & \textbf{Token} & \textbf{Rep} & \textbf{Cons.} & \textbf{Neuron} & \textbf{Layer} & \textbf{Token} & \textbf{Rep} & \textbf{Cons.} & \textbf{Neuron} & \textbf{Layer} & \textbf{Token} & \textbf{Rep} & \textbf{Cons.} \\
\midrule
\multirow{4}{*}{Jailbreak} & GCG & 0.994 & 0.873 & 0.430 & 0.980 & 0.877 & 0.994 & 0.842 & 0.563 & 0.980 & 0.779 & 0.977 & 0.923 & 0.601 & 0.992 & 0.971 \\
& AutoDAN & 0.994 & 0.858 & 0.833 & 0.984 & 0.939 & 0.994 & 0.712 & 0.868 & 0.951 & 0.746 & 0.977 & 0.864 & 0.798 & 0.988 & 0.977 \\
& AmpleGCG & 0.994 & 0.912 & 0.455 & 0.990 & 0.863 & 0.997 & 0.838 & 0.608 & 0.989 & 0.776 & 0.983 & 0.804 & 0.800 & 0.946 & 0.844 \\
& PAIR & 0.994 & 0.887 & 0.881 & 0.986 & 0.951 & 0.997 & 0.715 & 0.850 & 0.966 & 0.763 & 0.983 & 0.864 & 0.781 & 0.988 & 0.978 \\
\midrule
Hallucination & TruthfulQA & 0.526 & 0.661 & 0.642 & 0.476 & 0.671 & 0.524 & 0.649 & 0.698 & 0.591 & 0.697 & 0.523 & 0.670 & 0.698 & 0.598 & 0.696 \\
\midrule
\multirow{4}{*}{Backdoor} & Badnet & 0.980 & 0.823 & 0.671 & 0.961 & 0.801 & 0.947 & 0.863 & 0.556 & 0.992 & 0.755 & 0.944 & 0.979 & 0.909 & 0.983 & 0.871 \\
& CTBA & 0.947 & 0.754 & 0.671 & 0.952 & 0.744 & 0.947 & 0.834 & 0.620 & 0.947 & 0.704 & 0.947 & 0.730 & 0.634 & 0.980 & 0.709 \\
& MTBA & 0.954 & 0.646 & 0.663 & 0.923 & 0.667 & 0.951 & 0.672 & 0.667 & 0.968 & 0.686 & 0.947 & 0.896 & 0.760 & 0.992 & 0.668 \\
& Sleeper & 0.952 & 0.785 & 0.712 & 0.953 & 0.773 & 0.945 & 0.838 & 0.641 & 0.926 & 0.624 & 0.939 & 0.843 & 0.564 & 0.967 & 0.710 \\
\midrule
\multirow{3}{*}{Fairness} & Toxicity & 0.990 & 0.802 & 0.681 & 0.980 & 0.858 & 0.996 & 0.817 & 0.637 & 0.967 & 0.800 & 0.994 & 0.889 & 0.633 & 0.976 & 0.926 \\
& Sexually Explicit & 1.000 & 0.819 & 0.661 & 0.974 & 0.876 & 1.000 & 0.816 & 0.596 & 0.978 & 0.823 & 0.998 & 0.892 & 0.632 & 0.979 & 0.910 \\
& Severe Toxicity & 0.997 & 0.855 & 0.714 & 0.983 & 0.890 & 0.998 & 0.820 & 0.607 & 0.973 & 0.798 & 0.997 & 0.896 & 0.607 & 0.988 & 0.926 \\
\bottomrule
\end{tabular}
}
\end{table*}

\begin{table*}[t]
\centering
\caption{Detection performance (DSR) across jailbreak, hallucination, backdoor, and fairness tasks using different analysis on three LLMs.}
\label{tab:safety_eval_dsr}
\resizebox{\textwidth}{!}{
\begin{tabular}{ll|ccccc|ccccc|ccccc}
\toprule
\textbf{Task} & \textbf{Benchmark} & \multicolumn{5}{c|}{\textbf{LLaMA2}} & \multicolumn{5}{c|}{\textbf{Qwen}} & \multicolumn{5}{c}{\textbf{LLaMA3}} \\
\cmidrule(lr){3-7} \cmidrule(lr){8-12} \cmidrule(lr){13-17}
& & \textbf{Neuron} & \textbf{Layer} & \textbf{Token} & \textbf{Rep} & \textbf{Cons.} & \textbf{Neuron} & \textbf{Layer} & \textbf{Token} & \textbf{Rep} & \textbf{Cons.} & \textbf{Neuron} & \textbf{Layer} & \textbf{Token} & \textbf{Rep} & \textbf{Cons.} \\
\midrule
\multirow{4}{*}{Jailbreak} & GCG & 100.0\% & 94.4\% & 59.6\% & 98.4\% & 86.8\% & 100.0\% & 91.6\% & 47.2\% & 98.4\% & 72.9\% & 100.0\% & 96.0\% & 60.2\% & 100.0\% & 97.1\% \\
& AutoDAN & 100.0\% & 91.6\% & 83.6\% & 99.2\% & 93.9\% & 100.0\% & 69.6\% & 92.4\% & 92.8\% & 72.9\% & 100.0\% & 85.2\% & 77.1\% & 99.2\% & 97.7\% \\
& AmpleGCG & 100.0\% & 94.2\% & 61.2\% & 99.4\% & 85.5\% & 100.0\% & 90.0\% & 53.4\% & 99.0\% & 72.6\% & 100.0\% & 72.8\% & 77.2\% & 90.8\% & 86.2\% \\
& PAIR & 100.0\% & 89.6\% & 90.4\% & 98.6\% & 95.0\% & 100.0\% & 69.4\% & 90.4\% & 94.6\% & 74.4\% & 100.0\% & 82.4\% & 75.0\% & 98.8\% & 97.8\% \\
\midrule
Hallucination & TruthfulQA & 64.8\% & 87.7\% & 78.5\% & 54.8\% & 52.8\% & 64.4\% & 73.1\% & 100.0\% & 73.5\% & 53.5\% & 63.0\% & 84.9\% & 100.0\% & 74.4\% & 53.4\% \\
\midrule
\multirow{4}{*}{Backdoor} & Badnet & 98.6\% & 81.6\% & 64.0\% & 95.2\% & 81.2\% & 100\% & 85.4\% & 53.0\% & 98.8\% & 75.2\% & 100\% & 98.0\% & 90.4\% & 97.6\% & 86.2\% \\
& CTBA & 100\% & 73.4\% & 64.0\% & 96.0\% & 73.4\% & 100\% & 81.8\% & 60.8\% & 92.0\% & 60.4\% & 100\% & 71.4\% & 60.2\% & 99.2\% & 60.0\% \\
& MTBA & 100\% & 66.0\% & 62.6\% & 94.4\% & 50.0\% & 100\% & 67.0\% & 50.0\% & 95.2\% & 56.8\% & 100\% & 89.6\% & 72.6\% & 99.6\% & 50.2\% \\
& Sleeper & 100\% & 76.2\% & 66.4\% & 96.6\% & 70.8\% & 100\% & 83.8\% & 62.4\% & 92.0\% & 65.0\% & 99.2\% & 84.2\% & 57.8\% & 94.0\% & 59.8\% \\
\midrule
\multirow{3}{*}{Fairness} & Toxicity & 100.0\% & 73.6\% & 75.6\% & 98.4\% & 85.6\% & 100.0\% & 88.4\% & 62.0\% & 100.0\% & 77.7\% & 100.0\% & 86.4\% & 76.8\% & 96.0\% & 92.6\% \\
& Sexually Explicit & 100.0\% & 77.0\% & 72.0\% & 96.0\% & 87.4\% & 100.0\% & 87.0\% & 54.2\% & 100.0\% & 80.6\% & 100.0\% & 87.0\% & 75.0\% & 95.8\% & 91.1\% \\
& Severe Toxicity & 100.0\% & 84.6\% & 80.8\% & 98.0\% & 88.5\% & 100.0\% & 87.4\% & 58.0\% & 100.0\% & 77.8\% & 100.0\% & 88.0\% & 71.0\% & 98.0\% & 92.5\% \\
\bottomrule
\end{tabular}
}
\end{table*}
\section{Evaluation: Misbehavior Detection} \label{sec: rq3}
Having demonstrated in Section~\ref{sec: rq1} and Section~\ref{sec: rq2} that our multi-level framework effectively identifies critical safety components whose manipulation alters model behavior, we now examine whether the causal signals extracted from this framework can be harnessed to build novel detection systems for identifying safety-critical behaviors. Such systems would strengthen model defenses against emerging threats. While prior work has explored causality-guided jailbreak detection, as reviewed in Section~\ref{sec: background}, our investigation extends this line of research by systematically analyzing four levels of causal signals across multiple categories of model misbehavior, as detailed below.

\subsection{Misbehavior Detection}
In Section~\ref{sec: causal_framework}, we defined various metrics for quantifying causal correlations of specific components through intervention-based analysis. In this experiment, we investigate whether these causal metrics can effectively distinguish between benign prompts and adversarial prompts across different safety-critical tasks. 
Unlike the previous intervention evaluation that relied on safety judgment outcomes, which incur substantial computational costs and may not provide accurate assessments for certain safety-critical tasks, we now construct detection features by measuring the logits of the first generated token before and after intervention, providing an efficient metric to evaluate behavioral changes at token, layer, neuron, and representation levels derived from our causality analysis framework.

\textit{Token-Level Detection.} For token-level detection, we employ the ACE metric defined in Section~\ref{sec: causal_framework} but adopt a simplified comparison mechanism to avoid dependency on safety judgments. Specifically, for each token $x_i$ in a prompt $\mathbf{x} = (x_1, x_2, \ldots, x_n)$, we compute ACE based on the logits of the first generated token:
\begin{equation}
\text{ACE}_{x_i} = \left| \text{logits}_1(\mathbf{x}) - \text{logits}_1\left(\mathbf{x} , \middle| , \text{do}(x_i = \text{[PAD]})\right) \right|,
\end{equation}
where $\text{logits}_1(\mathbf{x})$ denotes the logit vector for the first generated token under the original prompt $\mathbf{x}$, and $\text{logits}_1(\mathbf{x} , | , \text{do}(x_i = \text{[PAD]}))$ represents the same logit vector after intervening on token $x_i$.  Since prompts vary in length, we cannot directly use raw ACE vectors as features. Instead, we extract features comprising the original first-token logit and statistical properties of ACE values across the entire sequence: mean, standard deviation, kurtosis, range, and skewness. These features are then fed into a multi-layer MLP classifier to distinguish between benign and adversarial prompts.

\textit{Layer-Level Detection.} For layer-level detection, we measure the causal effect of layer removal through first-token logit perturbations similar to token-level:
\begin{equation}
\text{ACE}_{\ell, n} = \left| \text{logits}_1(\mathbf{x}) - \text{logits}_1\left(\mathbf{x} , \middle| , \text{do}(\text{remove } [\ell:\ell+n])\right) \right|,
\end{equation}
where $\ell$ denotes the starting layer index and $n$ represents the span length. We conduct two types of interventions: single-layer removal ($n=1$) and three-consecutive-layer removal ($n=3$), recording ACE values for each configuration. 
The complete set of ACE values across all layer positions forms a feature vector that is input to MLP classifier for classification between benign and adversarial prompts.

\textit{Neuron-Level Detection.} For neuron-level detection, we leverage the logistic regression classifiers already trained to predict safety outcomes from hidden states at each layer. 
Since these layer-specific classifiers produce probability scores $p^{(\ell)} \in [0, 1]$ representing the likelihood of generating harmful content, 
we aggregate the prediction scores across all layers $\ell \in {1, 2, \ldots, L}$ to form a feature vector $\mathbf{p} = (p^{(1)}, p^{(2)}, \ldots, p^{(L)})$. 
This vector is then used as input to train an MLP classifier that distinguishes between benign and adversarial prompts based on the layer-wise safety prediction patterns.

\textit{Representation-Level Detection.} For representation-level detection, we compute geometric distance features across all transformer layers. 
We first apply PCA dimensionality reduction to the hidden states from both benign and harmful training data at each layer $\ell$, projecting them into a lower-dimensional space. 
We then compute cluster centroids $\mathbf{c}_{\text{benign}}^{(\ell)}$ and $\mathbf{c}_{\text{harmful}}^{(\ell)}$ for the benign and harmful data distributions at each layer. 
For each test prompt $\mathbf{x}$, we extract its hidden state $\mathbf{h}^{(\ell)}(\mathbf{x})$ at every layer, project it into the corresponding PCA space to obtain $\mathbf{h}_{\text{PCA}}^{(\ell)}(\mathbf{x})$, and compute distance-based features:
\begin{equation}
\begin{split}
d_{\text{benign}}^{(\ell)} &= \left\| \mathbf{h}_{\text{PCA}}^{(\ell)}(\mathbf{x}) - \mathbf{c}_{\text{benign}}^{(\ell)} \right\|_2, \\
d_{\text{harmful}}^{(\ell)} &= \left\| \mathbf{h}_{\text{PCA}}^{(\ell)}(\mathbf{x}) - \mathbf{c}_{\text{harmful}}^{(\ell)} \right\|_2.
\end{split}
\end{equation}
We aggregate these layer-wise distance features along with their relative ratios $d_{\text{benign}}^{(\ell)} / d_{\text{harmful}}^{(\ell)}$ across all layers to form a comprehensive feature vector, which is used as input to train a MLP classifier for prompt classification.

\textit{Layer Consistency Detection.} Beyond the intervention-based detection features described above, we also explore layer consistency as an alternative detection approach, inspired by prior work~\cite{min2025crow,chen2024inside} that leverages the smoothness of layer transitions to identify model misbehavior.
For each test prompt $\mathbf{x}$, we extract hidden states $\mathbf{h}^{(\ell)}(\mathbf{x})$ at every layer $\ell \in \{0, 1, \ldots, L-1\}$, where $L$ is the total number of layers.
At each layer, we aggregate the token-level representations by computing their mean:
\begin{equation}
\bar{\mathbf{h}}^{(\ell)}(\mathbf{x}) = \frac{1}{T} \sum_{t=1}^{T} \mathbf{h}_t^{(\ell)}(\mathbf{x}),
\end{equation}
where $T$ is the sequence length and $\mathbf{h}_t^{(\ell)}(\mathbf{x})$ denotes the hidden state of the $t$-th token at layer $\ell$.
We then measure the coherence between consecutive layers by computing cosine similarity:
\begin{equation}
s^{(\ell)}(\mathbf{x}) = \frac{\bar{\mathbf{h}}^{(\ell)}(\mathbf{x}) \cdot \bar{\mathbf{h}}^{(\ell+1)}(\mathbf{x})}{\left\| \bar{\mathbf{h}}^{(\ell)}(\mathbf{x}) \right\|_2 \left\| \bar{\mathbf{h}}^{(\ell+1)}(\mathbf{x}) \right\|_2}, \quad \ell \in \{0, 1, \ldots, L-2\}.
\end{equation}
The resulting layer consistency feature vector $\mathbf{s}(\mathbf{x}) = [s^{(0)}(\mathbf{x}), s^{(1)}(\mathbf{x}), \ldots, s^{(L-2)}(\mathbf{x})] \in \mathbb{R}^{L-1}$ captures the smoothness of representation transitions throughout the model. Similar to other detection approaches, this feature vector is fed into a classifier (Logistic Regression or MLP) to distinguish between benign and adversarial prompts.

All MLP classifiers are implemented with two hidden layers of 128 and 64 neurons respectively.

\subsection{Experimental Setup}
We evaluate our causal-signal based detection methods across four categories of safety-critical behaviors, i.e., Jailbreak, Hallucination, Backdoor, and Fairness, where each category corresponds to a distinct model failure mode and is assessed using established benchmark datasets and attack frameworks. For every category, we construct a mixed dataset containing both benign prompts and misbehavior-inducing prompts. Benign prompts are sampled from the Alpaca dataset~\cite{alpaca}, and for each task we ensure a balanced number of benign and adversarial examples in both the training and testing splits.
\begin{itemize}
    \item For jailbreak detection, we consider four representative jailbreak attacks: AutoDAN~\cite{AutoDAN24}, PAIR~\cite{PAIR2023Chao}, and GCG~\cite{GCG2023Zou} and AmpleGCG~\cite{amplegcg2024}. For each method, we generate 500 adversarial prompts using AdvBench~\cite{GCG2023Zou}. We train an MLP classifier on 50\% of the GCG and AutoDAN data and evaluate on all four attack types to measure both classification accuracy and cross-attack transferability.

    \item For hallucination detection, we use TruthfulQA~\cite{lin-etal-2022-truthfulqa}, which includes prompts that intentionally provoke hallucinations. The classifier is trained on 50\% of the prompts and tested on the remaining 50\%.

    \item For backdoor detection, we study three backdoor attack families—BadNets~\cite{Gu2017BadNetsIV}, CTBA~\cite{Huang2023CompositeBA}, MTBA~\cite{Li2024MultiTriggerBA}, and Sleeper~\cite{Hubinger2024SleeperAT}. We finetune LLMs using the corresponding poisoned datasets, generate 500 trigger-embedded prompts for each method, and perform detection on the resulting backdoored models. As before, 50\% of the prompts are used for training and 50\% for testing.

    \item For discrimination detection, we evaluate fairness-related misbehavior using RealToxicityPrompts~\cite{Gehman2020RealToxicityPromptsEN}, focusing on three adversarial subsets: \textit{adversarial\_severe\_toxicity}, \textit{adversarial\_sexually\_explicit}, and \textit{adversarial\_toxicity}. We train the classifier on 50\% of the \textit{adversarial\_toxicity} subset and test on the remaining samples.
\end{itemize}

We evaluate our detection methods using two metrics: Detection Success Rate (DSR), computed as the ratio of detected adversarial prompts to total adversarial prompts, and F1-score to provide a balanced assessment of detection performance.

\begin{table}[!t]
\centering
\caption{Hallucination detection using multi-level feature fusion, showing F1 scores and DSR (in parentheses) across LLaMA2-7B, Qwen2.5-7B, and LLaMA3.1-8B.}
\label{tab:multi_level}
\resizebox{0.45\textwidth}{!}{%
\begin{tabular}{lccc}
\hline
& LLaMA2-7b & Qwen-7B & LLaMA3-8B  \\
\hline
Token + Layer & 0.672 (91.4\%) & 0.564 (57.8\%) & 0.663 (72.2\%)  \\
Rep + Neuron & 0.995 (99.4\%) & 0.981 (97.8\%) & 0.753 (61.2\%)  \\
All Combined & 0.987 (100\%) & 0.956 (99.4\%) & 0.971 (97.0\%)  \\
\hline
\end{tabular}
}
\end{table}

\subsection{Experimental Results}
Based on the comprehensive evaluation results presented in Tables~\ref{tab:safety_eval_f1} and~\ref{tab:safety_eval_dsr}, our multi-level causality analysis framework demonstrates strong detection capabilities across diverse safety-critical tasks, with varying performance across different causal signals.

Neuron-level and representation-level methods based on hidden states achieved the best performance in jailbreak, backdoor, and fairness detection tasks. Neuron-level detection maintained F1 scores above 0.977 with 100\% DSR for jailbreak detection, and achieved near-perfect results (F1: 0.990-1.000, DSR: 100\%) for fairness tasks. Representation-level detection showed similarly robust performance, with F1 scores ranging from 0.946 to 0.992 for jailbreak detection and 0.952-0.961 for backdoor detection. Additionally, layer-level analysis significantly outperformed token-level analysis across most tasks. This is understandable since tokens vary in input length and can only provide statistical data. While changes to single tokens in tasks like GCG or backdoor may directly affect outcomes, the varying token input lengths means statistical values may fail to capture these changes. For instance, GCG detection via token-level analysis achieved only 0.430-0.608 F1 scores, compared to 0.842-0.923 for layer-level methods.

For the hallucination task, although token-level analysis achieved 100\% DSR on both Qwen and Llama3,
the low F1 scores (0.642-0.698) suggest this may indicate oversensitivity rather than accurate detection. 
Notably, F1 scores for hallucination detection remained below 0.7 across all methods, 
suggesting that new approaches are needed for this task. 

To address this limitation, we explore combining causal signals from different levels rather than relying on a single feature type. As demonstrated in Table~\ref{tab:multi_level}, multi-level feature combinations significantly improve detection performance. While token and layer features alone achieve moderate results (F1: 0.564-0.672), combining representation and neuron features yields substantial improvements (F1: 0.753-0.995). The combination of all feature levels achieves the best performance across all models, with F1 scores ranging from 0.956 to 0.987 and DSR from 97.0\% to 100\%, demonstrating that integrating complementary causal signals from multiple levels provides more robust and accurate hallucination detection.

\begin{table}[!t]
\centering
\caption{Average detection time per input (seconds) across four causality analysis levels on three LLMs.}
\label{tab:detection_time}
\resizebox{0.45\textwidth}{!}{%
\begin{tabular}{lccc}
\hline
Model & Llama2-7b & Qwen-7B & Llama3-8B  \\
\hline
Neuron-Level & 0.12 & 0.12 & 0.14  \\
Layer-Level & 2.11 & 1.92 & 2.85  \\
Token-Level & 2.87 & 4.08 & 4.37  \\
Rep-Level & 0.07 & 0.08 & 0.10  \\
Layer-Consistency & 0.05 & 0.05 & 0.09 \\
\hline
\end{tabular}
}
\end{table}

Table~\ref{tab:detection_time} demonstrates substantial efficiency differences across analysis levels. Neuron-level and representation-level methods require only 0.07-0.14 seconds per input through single-pass inference based on hidden states from all layers. In contrast, token-level and layer-level methods require multiple inference runs (1.92-4.37 seconds per input). Token-level analysis is particularly costly (2.87-4.37 seconds) as interventions scale linearly with unpredictable input length, while layer-level analysis incurs moderate costs (1.92-2.85 seconds) with a fixed number of interventions determined by model architecture.

Based on both detection performance and computational efficiency, neuron-level and representation-level detection methods emerge as the most practical approaches for real-world deployment. 
These methods consistently achieve superior accuracy across jailbreak, backdoor, and fairness tasks while maintaining the lowest computational overhead through single-pass inference. 

\section{Conclusion}
In this work, we present a comprehensive framework for causality analysis in understanding and controlling model behavior. By systematically examining how input token, neurons, layers of neurons, and internal representations causally influence outputs, we demonstrate that both attacks and defenses can be framed as interventions informed by causal insights. Our review shows that recent attack methods are becoming increasingly effective and require less prior knowledge of the target model, making them more practical and emphasizing the urgent need for robust, causality-informed defenses.

From a defense perspective, causality analysis enables precise interventions, such as targeted model editing, unlearning, and layer-specific modifications, which can mitigate unsafe behaviors while preserving beneficial capabilities. This paradigm highlights that understanding the causal mechanisms underlying model behavior is not only critical for explaining vulnerabilities but also for designing principled and effective safety strategies.

Looking forward, several avenues for future research are particularly promising. One key direction is the discovery of additional causally relevant components or signals within LLMs that may reveal hidden vulnerabilities or latent safety mechanisms, such as internal temporal and spatial consistency. Systematically identifying these components across different architectures and training regimes could deepen our understanding of where models fail and how to intervene. Another important area is conducting rigorous, comparative studies of existing causality-based approaches to evaluate their relative effectiveness, efficiency, and generalizability. Such benchmarking would provide clearer guidance for practitioners seeking to deploy safe models in real-world settings.

Finally, there is significant scope for developing more advanced causality-informed defenses. This could include methods that combine token-level, layer-level, neuron-level, and representation-level interventions, or approaches that dynamically adapt safety mechanisms based on ongoing causal analysis during inference. By pursuing these directions, the research community can move toward a more holistic understanding of model vulnerabilities and defenses, ultimately enabling LLMs that are both powerful and reliably safe.

\newpage
\bibliographystyle{ieee_fullname}
\bibliography{sample}


\section{Appendix}
\label{sec: appendix}

\end{document}